# Intermediate Resistive State in Wafer-Scale MoS$_2$ Memristors through Lateral Silver Filament Growth for Artificial Synapse Applications

Yuan Fa[1,2], Milan Buttberg[3], Ke Ran[1,4,5], Rana Walied Ahmad[6], Dennis Braun[2], Lukas Völkel[2], Jimin Lee[2], Sofía Cruces[2], Bart Macco[7], Bárbara Canto[1], Holger Lerch[1], Thorsten Wahlbrink[1], Holger Kalisch[8], Michael Heuken[8,9], Andrei Vescan[8], Joachim Mayer[4,5], Zhenxing Wang[1], Ilia Valov[6,10], Stephan Menzel[6], and Max C. Lemme[1,2]*

1 AMO GmbH, Advanced Microelectronic Center Aachen, Otto-Blumenthal-Str. 25, 52074 Aachen, Germany

2 Chair of Electronic Devices (ELD), RWTH Aachen University, Otto-Blumenthal-Str. 25, 52074 Aachen, Germany

3 Institute of Materials in Electrical Engineering and Information Technology II (IWE2), RWTH Aachen University, Sommerfeldstraße 24, 52074 Aachen, Germany

4 Central Facility for Electron Microscopy (GFE), RWTH Aachen University, Ahornstr. 55, 52074 Aachen, Germany

5 Ernst Ruska-Centre for Microscopy and Spectroscopy with Electrons (ER-C 2), Forschungszentrum Jülich GmbH, Wilhelm-Johnen-Str., 52425 Jülich, Germany.

6 Peter Grünberg Institute 7 (PGI 7) and JARA-FIT, Forschungszentrum Jülich GmbH, Wilhelm-Johnen-Straße, 52428 Jülich, Germany

7 Department of Applied Physics and Science Education, Eindhoven University of Technology, 5600 MB Eindhoven, The Netherlands

8 Compound Semiconductor Technology (CST), RWTH Aachen University, Sommerfeldstr. 18, 52074 Aachen, Germany

9 AIXTRON SE, Dornkaulstr. 2, 52134 Herzogenrath, Germany

10 "Acad. Evgeni Budevski" IEE-BAS, Bulgarian Academy of Sciences (BAS), Acad. G. Bonchev Str, Block 10, Sofia 1113, Bulgaria

Email: max.lemme@rwth-aachen.de

# Abstract

Memristors based on two-dimensional materials (2DMs) have garnered significant attention due to their fast resistive switching (RS) behavior and atomic-level thickness, which enables low power consumption, making them promising candidates for neuromorphic computing. Among these, memristors based on molybdenum disulfide ($MoS_2$) have been extensively studied. Their RS has been attributed to the formation and rupture of conductive filaments (CFs). However, the underlying mechanism of filament formation remains underexplored, and the inherently stochastic nature of RS leads to high variability and limited reproducibility. Additionally, the lack of scalable fabrication techniques for 2DM-based memristors restricts their integration into standard semiconductor technology. Here, we demonstrate memristors based on metal–organic chemical vapor-deposited $MoS_2$ on the wafer-scale. Our devices exhibit volatile and nonvolatile RS behavior, tunable by modulating the current compliance. Notably, we observe stable RS characteristics in an intermediate resistive state (IRS), featuring set and reset voltages within ±1 V, an endurance exceeding 2500 cycles in direct current mode, and a state retention over $10^6$ s. The experimental data, complemented with simulations, suggest that the IRS originates from the lateral growth of the CF within the $MoS_2$ layer. Furthermore, the devices successfully emulate synaptic plasticity with current responses on the microsecond timescale, highlighting their potential for large-scale integration in neuromorphic computing architectures.

# Introduction

The rapid advancement of artificial intelligence (AI) and big data is profoundly impacting various aspects of human life. As AI technologies are increasingly adopted for tasks such as text and image generation, the energy consumption associated with data transfer in conventional von Neumann chip architectures has emerged as a critical challenge.[1–4] Brain-inspired neuromorphic computing and electronic devices tailored for AI applications offer beyond-von Neumann architectures.[5,6] Among these devices, vertical electrochemical metallization (ECM) memristors have attracted attention because of their simple metal–insulator–metal architecture and high integration density.[7–10] ECM memristors exhibit resistive switching (RS) behavior governed by a filamentary switching mechanism, in which the application of an electric field induces the formation and rupture of conductive filaments (CFs) through ion migration from the active electrodes, enabling the reversible modulation of resistive states with low operating voltages.[11,12] Furthermore, their Computing-In-Memory capabilities help bridge the gap between memory and processing, enhancing their suitability for neuromorphic computing applications.[11,13,14]

Two-dimensional materials (2DMs) have emerged as promising alternatives, and their RS parameters, such as retention and endurance, are comparable to those of conventional oxide-based memristors.[15] Moreover, their atomic-scale thickness,[16–18] even down to a single atomic layer that cannot be realized in oxide-based counterparts, offers the potential for high switching speeds and low-power synaptic applications.[11,19,20] Among various 2DMs, molybdenum disulfide ($MoS_2$) has attracted considerable attention owing to its favorable RS characteristics and compatibility with device miniaturization.[21–28] Recently, monolayer $MoS_2$-based memristors, often referred to as "atomristors," have enabled subnanometer-thick active layers, offering scalability and structural controllability.[22,27] For multilayer MoS$_2$ memristors, Huang et al. demonstrated that increasing the layer thickness consistently improved the device yield and endurance.[29] Furthermore, the integration of MoS$_2$ memristors into a two-transistor one-resistor (2T-1R) architecture represents a significant step toward the development of nonvolatile memory systems.[25] Despite these advances, key challenges remain. One of them is the scalable fabrication of high-performance MoS$_2$ memristors. Another critical issue is the inherently stochastic nature of RS behavior, particularly in vertical filament-type MoS$_2$ memristors.[7] This randomness primarily originates from the uncontrolled formation and rupture of CFs driven by ion migration. In situ high-resolution transmission electron microscopy studies have captured filament evolution in oxides[30] and in van der Waals (vdW) gaps in lateral 2D MoS$_2$ memristors.[31] However, such experiments are not yet available for vertical 2D memristors. Kaniselvan et al. systematically reviewed RS mechanisms in filamentary-type multilayer 2DM-based memristors; however, the specific role of the vdW gaps for vertical filament formation remains largely unexplored.[7]

In this work, we report wafer-scale Ag/MoS$_2$/Pd memristors exhibiting RS with an intermediate resistive state (IRS). MoS$_2$ grown via metal–organic chemical vapor deposition (MOCVD) serves as the switching medium, with silver (Ag) and palladium (Pd) as the top and bottom electrodes, respectively. Across the wafer, the IRS consistently emerges during current–voltage (*I-V*) sweeps, demonstrating endurance over 2500 cycles and retention exceeding $10^6$ s. To probe the underlying switching mechanism, we employed a combination of energy-dispersive X-ray (EDXS) mapping, mathematical analysis, temperature-dependent measurements, and analytical modeling. The experimental observation of the IRS is further supported by a physics-based continuous model complemented by a compact model. These link the IRS behavior to the lateral growth of Ag filaments perpendicular to the main vertical filaments, enabled by the vdW gaps. Under pulsed voltage stimuli, the devices exhibit synaptic plasticity, underscoring their potential for neuromorphic computing applications.

## Results

### Material and Device Characterization

We used MoS$_2$ grown by MOCVD on a 2-inch sapphire substrate for our experiments. The MoS$_2$ was first characterized via atomic force microscopy (AFM) on the growth substrate. A scan 2 × 2 µm$^2$ area revealed triangular domains, characteristic of MoS$_2$ grain boundaries and indicative of its polycrystalline nature (Figure 1a). The root-mean-square (RMS) surface roughness was approximately 1 nm. We further performed an AFM scan across a 5 × 10 µm$^2$ region with a scratch and measured a MoS$_2$ film thickness of approximately 4.3 nm (Figure 1b), in line with multilayer MoS$_2$. The chemical composition and stoichiometry of the as-grown MoS$_2$ were examined via X-ray photoelectron spectroscopy (XPS). As shown in Figures 1c and 1d, the S 2p$_{3/2}$ and S 2p$_{1/2}$ peaks appear at 162.6 eV and 163.8 eV, respectively, whereas the Mo 3d$_{5/2}$ and Mo 3d$_{3/2}$ peaks are located at 229.7 eV and 232.9 eV. These values are consistent with those reported in the literature for the 2H-phase of MoS$_2$.[32,33] The stoichiometric ratio of S to Mo is close to 2, and the sharp peaks indicate a single oxidation state and a low defect concentration, confirming the high quality of the MOCVD MoS$_2$.[21]

We then fabricated vertical memristors with Ag and Pd electrodes. A schematic illustration of one such Ag/MoS$_2$/Pd memristor is shown in Figure 1c, while the inset shows top-view microscopy images of a processed 6-inch wafer and a representative crosspoint device featuring an active area of 16 µm². The bottom electrodes (BEs), composed of 45 nm Pd with a 5 nm titanium (Ti) adhesion layer, were defined via stepper lithography, followed by electron-beam evaporation and a lift-off process. The MOCVD MoS$_2$ on the sapphire wafer was subsequently transferred onto the device substrate via a poly(methyl methacrylate) (PMMA)-assisted wet-transfer technique. Raman spectroscopy (see the Supporting Information (SI) Figure S1) verified the structural preservation of MoS$_2$ during the transfer process, as evidenced by the unchanged positions of the characteristic $E^1_{2g}$ (382.5 cm$^{-1}$) and $A_{1g}$ (407.2 cm$^{-1}$) modes. The transferred MoS$_2$ was patterned using stepper lithography and reactive ion etching (RIE) with a CF$_4$/O$_2$ plasma. 50 nm of Ag was deposited as the TEs, followed by a lift-off process. As a result, the MoS$_2$ layer, serving as the active RS medium, was sandwiched between the BEs and TEs. Additional details regarding the fabrication process are provided in the Experimental Section.

A focused ion beam (FIB) lamella was prepared from a memristor before any electrical measurements for high-resolution transmission electron microscopy (HRTEM). The HRTEM image in Figure 1f reveals the cross-sectional structure of the device comprising Ag, MoS$_2$, and Pd layers. Within this local region, 6-layer MoS$_2$ can be identified with a uniform interlayer spacing of ~0.65 nm, which is in agreement with the literature values for 2D MoS$_2$.[34] In Figure 1g, the composite map of Ag, sulfur (S), and Pd based on EDXS elemental mapping was recorded from the same lamella, showing good agreement with our device design (see the complete mapping results in SI Figure S2). We particularly observed the carbon (C) EDX signal, as certain amounts of residual PMMA or other resist may be expected at the Ag/MoS$_2$ interface due to the transfer and lithography processes. These have been shown to exhibit RS behaviors very similar to those reported for 2DMs and could compromise the interpretation of RS switching in 2DM memristors.[35] In the present case, the C map in SI Figure S3c shows only background-level signals, leading us to rule out such a parasitic effect.

## Electrical Performance in Direct Current (DC) Mode

We conducted *I–V* measurements in DC mode to evaluate the switching performance of our wafer-scale MOCVD-grown $MoS_2$-based memristors. In all the measurements, a voltage was applied to the Ag TE, while the Pd BE was grounded. Figure 2a displays 30 reproducible unipolar threshold switching (TS) curves for the positive voltage polarity under three different current compliance ($I_{cc}$) levels: 2.5 µA, 5.0 µA, and 7.5 µA, sequentially conducted in that order. Upon sweeping the voltage from 0 V to approximately 0.35 V, devices in the high resistive state (HRS) transitioned abruptly to a volatile low resistive state (vLRS) at a distinct threshold voltage ($V_t$). When the voltage was swept back to 0 V, the device returned to the HRS, resulting in a complete anticlockwise TS hysteresis loop. Notably, $V_t$ decreased progressively with increasing $I_{cc}$. We attribute this trend to the sequential nature of the measurements rather than the $I_{cc}$ itself. This suggests that the CF pathways formed during earlier cycles may facilitate easier CF reformation, thus lowering $V_t$ in subsequent measurements. To rule out the influence of $I_{cc}$ on the observed $V_t$ trend, additional measurements were performed on a separate device using a randomized $I_{cc}$ sequence (SI Figure S3). These results confirmed that $V_t$ is independent of $I_{cc}$, supporting our interpretation. Figure 2b presents a histogram of $V_t$ values at each $I_{cc}$, fitted with Gaussian distributions. The corresponding cumulative distribution functions (CDFs) are shown in SI Figure S4. The narrow distribution, with a standard deviation of ~2% highlights the low variability of the TS behavior (see Table S1 for detailed statistics). We also performed *I-V* sweeps on control devices fabricated identically but without the $MoS_2$ layer. These devices did not exhibit RS behavior (SI Figure S5), confirming the essential role of $MoS_2$ in the switching process.

We subsequently increased the $I_{cc}$ during bipolar *I–V* measurements on the same device and observed forming-free nonvolatile switching (NVS). When $I_{cc}$ was initially set to 1 mA, the device exhibited unstable NVS behavior with chaotic *I–V* curves (SI Figure S6; arrows indicate the voltage sweep direction). Upon repeated cycling, the HRS progressively degraded, narrowing the RS window. We attribute this phenomenon to the simultaneous growth of CFs in both vertical and lateral directions, as further supported by the simulation results, highlighting the necessity of optimizing the $I_{cc}$ for stable performance.

To achieve more controlled NVS, we tested a pristine device beginning with a lower $I_{cc}$ of 100 μA. As shown by the 35 darker *I–V* curves in Figure 2c, the device transitioned from the HRS to an intermediate resistive state (IRS) at a set voltage ($V_{set}$) of approximately 0.35 V, where the current abruptly rose to the $I_{cc}$ limit. The IRS was retained after the voltage returned to 0 V. Upon applying a negative voltage, the device switched back to the HRS at a reset voltage ($V_{reset}$) of approximately 0.3 V. Note that we define this state as an IRS because our devices can subsequently switch into a true low resistive state (LRS) with significantly lower resistance, and the IRS is both electrically and mechanistically distinct from the LRS. After 35 NVS cycles between the HRS and the IRS, the device could no longer be reset to the HRS (SI Figure S7a). Attempts to apply a higher negative bias to reset to the HRS in another device under the same conditions resulted in permanent shorting (SI Figure S7b). We attribute this non-resettable failure to desulfurization of the $MoS_2$ and the development of a critical gap size between the TE and the filament, accompanied by suppressed filament retraction (see more details in the simulation section). Eventually, the accumulated remnant filament can no longer be ruptured by the applied reverse bias, leaving the device irreversibly set to the IRS. After stabilization in the IRS, we increased the $I_{cc}$ to 1 mA and observed further switching into the LRS. While the same measurement protocol was used, the RS between the IRS and the LRS was notably more gradual than the abrupt RS during the HRS ↔ IRS transitions, which we attributed to the lateral expansion of the Ag CF.

To quantitatively assess the switching uniformity, we plotted histograms of $V_{set}$ and $V_{reset}$ of 35 NVS cycles at 100 µA and 100 cycles at 1 mA (data from Figure 2c) in Figure 2d. Each histogram follows a Gaussian distribution. The corresponding CDFs are shown in SI Figure S8. Notably, both set and reset operations were achieved within ±1 V, meeting the voltage requirements for low-power memory applications.[36] In addition, $V_{set}$ exhibited a smaller standard deviation than $V_{reset}$ did, indicating higher cycle-to-cycle stability for the set process. The statistical details are provided in SI Table S2. Figure 2e shows the resistance values of the four distinct resistive states measured at a read voltage ($V_{read}$) of 0.1 V and plotted versus $I_{cc}$. The box plot distinguishes the resistance levels of each state, with mean values of 1.49±0.16 MΩ for the HRS, 2.41±0.75 kΩ for the IRS, and 0.31±0.10 kΩ for the LRS. The device performance was further examined through endurance and retention tests. Figure 2f shows stable NVS during the IRS ↔ LRS transitions of at least 2500 cycles at $V_{read}$ = 0.1 V, maintaining an on/off ratio of approximately 10. The clear separation between the IRS and LRS across cycles underscores the consistency and robustness of the switching behavior. The three resistance states demonstrated retention stability over $10^6$ s at $V_{read}$ = 0.1 V (Figure 2g), satisfying key requirements for nonvolatile memory applications.[37,38]

A total of 100 devices were randomly selected across the wafer to evaluate yield and device-to-device variability. Among the 100 devices tested, 98 exhibited reproducible RS during the IRS ↔

LRS transitions with an on/off ratio of approximately 10, reflecting a 98% yield and demonstrating the uniform and intrinsic nature of the IRS behavior in Ag TE-based memristors. We plotted the first NVS cycle in these 98 devices (Figure S9a), and statistical analysis of the first NVS cycle from 98 devices revealed low device-to-device variability, with $V_{set}$ = 0.20±0.05 V and $V_{reset}$ = -0.30±0.12 V (Figure S9b). These reproducible multistate switching characteristics support the feasibility of these devices as analog memory elements for neuromorphic computing architectures.[39]

## Investigation of Filamentary Switching

We examined $Ag^+$ ion migration and filament formation with different methods to further elucidate the RS mechanism. Figure 3a presents EDXS elemental line scan profiles of Ag, S, and Pd across the cross-section of a memristor in the IRS, acquired from a high-angle annular dark-field scanning transmission electron microscopy (HAADF-STEM) image of a lamella prepared for TEM analysis. The scan was conducted from the TE to the BE, and the inset elemental map illustrates the spatial distribution of the elements. In contrast to the line scan of a memristor in the pristine HRS (Figure S10), an Ag signal was detected near the surface of the BE and then returned to the background, as highlighted by the black dashed box in Figure 3a. This observation suggests $Ag^+$ ion migration during switching and the formation of a remnant filament after the reset process, supporting the proposed filamentary switching mechanism in the IRS. Furthermore, distinct Ag signal peaks were observed at positions corresponding to the minima of the S signals in the line scan, as indicated by the dashed lines. This spatial anti-correlation suggests that Ag atoms were likely located in the vdW gaps between the $MoS_2$ layers. These findings support the hypothesis that lateral filament growth is confined within the interlayer spaces of $MoS_2$, similar to filament growth observed in lateral MoS$_2$ memristors.[31,40] Further details on the lateral filament growth are provided in the simulation section.

Area-dependent measurements were performed on five cross-point memristors with varying electrode sizes. These devices exhibited both HRS ↔ IRS and IRS ↔ LRS switching under $I_{ccs}$ of 100 µA and 1 mA, respectively. The corresponding resistances were recorded over five I–V cycles. As shown in Figure 3b, the resistances in both the IRS and LRS are independent of the electrode area, indicating current conduction through CFs. In contrast, the HRS exhibits a strong area dependence, indicating that current flows through the entire active device area. This further corroborates the hypothesis of filamentary switching in our devices.[7,41]

We subsequently fitted the slopes of the IRS and LRS curves extracted from averaged I–V characteristics plotted on a double-logarithmic scale (SI Figure S11a). These averaged curves were calculated from the individual I–V sweeps shown in Figure 2c. As depicted in SI Figure S11b, both the IRS and LRS exhibit linear behavior with slopes of approximately 1 and coefficients of determination ($R^2$) > 0.999, indicating ohmic (metallic) behavior. These results further support the existence of a CF.

To elucidate the underlying conduction mechanisms, we performed temperature-dependent measurements on three devices in each of the LRS, IRS, and HRS states. Figures 3c-3e show the I–V curves of the LRS, IRS, and HRS recorded over a temperature range from 20 to 250 K. Notably, each state exhibited a distinct temperature dependence, as shown in Figure 3f: the LRS displayed an increase in resistance with increasing temperature, the HRS showed a decrease, and the IRS exhibited negligible temperature sensitivity. The LRS temperature-dependent I-V data revealed linear behavior with $R^2$ > 0.999 (see details in SI Table S3), indicating that ohmic conduction predominated.[42,43] The temperature dependence of the resistance in the LRS aligns with Matthiessen's rule:[44]

$$\rho(T) = \rho_0 + \rho_{\text{ph}}(T) \tag{1}$$

where $\rho_0$ is the residual resistivity and $\rho_{\text{ph}}$(T) arises from electron–phonon scattering. Accordingly, the temperature dependance of the resistance in metals can be described as follows:[45]

$$R(T) = R_0 + R_0 \cdot \alpha_m \cdot (T - T_0) \tag{2}$$

where $R_0$ is a baseline resistance at $T_0$ (often $T_0 \approx$ 0 K), and $\alpha_m$ is a material-specific temperature coefficient. Fitting the LRS data in Figure 3d yields $\alpha_m$ = 0.0034 K$^{-1}$, closely matching the reported value for Ag (0.0038 K$^{-1}$).[46] This strongly supports the formation of a metallic Ag filament as the dominant conduction path in the LRS. The weak temperature dependence of the resistances in the IRS suggests that tunneling currents govern the conduction process, as described by the Simmons model.[47] Further discussion on the IRS conduction mechanisms will be provided in the subsequent simulation section. For the HRS, fitting the temperature-dependent *I–V* characteristics suggests hopping conduction as the dominant conduction mechanism (see discussion in SI and SI Figure S12).

## Model-Assisted Interpretation of Switching Dynamics

We applied two recently published ECM cell models to analyze the *I-V* characteristics obtained from our experiments. The first is a one-dimensional (1D) compact model, and the second is a 2D axisymmetric continuum model. In the following, we summarize the key assumptions of each model along with the critical parameters required to simulate the *I-V* characteristics. The parameters used for the 1D compact and 2D continuum models are provided in SI Table S4 and SI

Table S5, respectively. We subsequently compare the outputs of both approaches, with particular emphasis on the discrepancies in filament evolution and switching behavior. Based on this comparative analysis, we propose a mechanism of filament formation and stabilization in our $MoS_2$-based ECM cells. The discussion highlights the distinct anisotropic behavior observed in layered vdW materials, in contrast to conventional isotropic ion-conducting systems. We then conclude by identifying the most favorable electrochemical reaction pathways that drive RS and filament formation.

## 1D Compact Model

An established 1D compact model for ECM cells was employed,[48–50] incorporating key phenomena such as redox reactions at the metal/$MoS_2$ interface, ion migration via a hopping mechanism, 1D vertical filament growth with a constant radius, electronic tunneling currents, and an $I_{cc}$ condition. Notably, the model assumes a fixed CF radius during growth, thereby excluding any lateral growth dynamics. Figure 4a displays the applied bias modeled with two consecutive triangular voltage sweeps with a distinct maximum $I_{cc}$. The first triangular sweep corresponds to the final sweep in the HRS ↔ IRS measurement sequence, after which the filament becomes non-resettable and transitions into the IRS ↔ LRS switching for higher current compliances. The second sweep corresponds to the subsequent IRS ↔ LRS switching cycles. Figure 4b shows the experimental I–V data alongside the simulation results of the 1D compact model shown as the gray solid line. Upon reaching a minimum physical gap of 0.07 nm between the electrode and the filament, further vertical filament growth is inhibited, resulting in the stabilization of the IRS. Subsequent LRS switching fails because of the critical gap size and the suppression of additional filament growth. Thus, this well-established model is suitable for describing the HRS ↔ IRS transition but not the IRS ↔ LRS transition, as observed in our MoS$_2$ memristors. Figure 4c shows the resistance trends during IRS and LRS switching for the experimental and the simulation data, respectively. Since the 1D model already reaches the minimum gap during IRS, no further filament growth occurs, resulting in an unchanged resistance.

## 2D Axisymmetric Continuum Model

We further applied a previously reported continuum model,[51,52] incorporating the key phenomena governing device operation, including redox reactions at the metal/MoS$_2$ interfaces, ion migration within MoS$_2$ via a hopping mechanism, tunneling currents, conductive filament formation using a moving mesh algorithm, tunnel barrier heating,[53] active electrode dissolution, and mechanically induced stress that constrains filament growth.[54,55] The stable, non-resettable IRS state was attributed to a remnant conductive filament, based on our experimental data. This filament was regarded as having a mixed origin, consisting of reduced Ag$^+$ ions from the active electrode (AE; here: the Ag TE) and Mo$^{(0)}$ generated via the desulfurization of MoS$_2$. The desulfurization process occurs over several set cycles and is assumed to be irreversible, resulting in the remnant filament.

As the Ag TE dissolves, a void domain forms, preventing further electrochemical activity. Consequently, vertical filament growth is inhibited, and lateral expansion ensues. This lateral expansion of the filament results in a linearly increasing current response. Figure 4b presents the simulation results of the continuum model as indicated by the light, medium, and dark red solid lines (labeled A-C) alongside the corresponding measurements. The simulation reproduced the final set cycle preceding the stable IRS, using a maximum voltage of 1 V, a rise time of 1.5 s, and an $I_{cc}$ of 100 µA. However, the model cannot yet account for the remnant filament. Hence, the reset process was omitted in the simulation. Instead, a stable, non-resettable filament was assumed, and the simulation proceeded with a subsequent set pulse featuring an increased $I_{cc}$ of 1 mA. In this configuration, the final HRS ↔ IRS cycle, prior to non-resettable filament stabilization with $I_{cc}$ = 100 µA, was simulated, followed by the IRS ↔ LRS cycle with $I_{cc}$ = 1 mA. In the HRS ↔ IRS transition regime, directed vertical filament growth across a gradually decreasing gap governs the switching dynamics, resulting in the characteristic exponential tunnel current response. The mismatch observed in the experimental HRS ↔ IRS sweeps, particularly the negative differential resistance during switching, is likely caused by the limited regulation speed of the Keithley 4200-SMU used during the measurements[56] and cannot be readily reproduced in the simulation without considerable complexity.

Nevertheless, the simulation captured the essential features of the final HRS ↔ IRS transition and reached the experimental IRS resistance level with good accuracy, providing a reliable basis for interpreting the subsequent transition into the IRS ↔ LRS regime. Here, lateral filament growth was assumed to dominate the switching dynamics. The simulations showed a gradual current increase that closely matched the experimental data, including its higher cycle-to-cycle variability than during the HRS ↔ IRS RS. In fact, the HRS ↔ IRS transitions exhibited almost no variability, indicating a stabilized filament configuration prior to lateral expansion. Within the IRS ↔ LRS regime, two characteristic current trends were observed: a purely gradual increase up to the compliance current $I_{cc}$ and a gradual rise followed by a sudden exponential surge. Both behaviors are successfully captured by the model through variation in the lateral mechanical stress. This is justified, as localized mechanical stress can vary in our experiments due to defects and dislocations within the $MoS_2$, as well as electrode interface inhomogeneities. High mechanical stress favors narrower filaments, which concentrate local heat and lead to an abrupt increase in current. In contrast, low mechanical stress promotes the formation of broader filaments, resulting in smoother, more gradual current evolution due to distributed heating over a larger volume. This dependency is consistent with the experimental trends, and the two characteristic curves in the IRS ↔ LRS characteristics are accurately reproduced within the modeling framework.

However, the final LRS value is overestimated or underestimated in the simulation. This discrepancy likely arises because the current model does not account for the formation of a fully ohmic contact between the filament and the TE, which is expected to form as the remaining simulated gap narrows to physically meaningful dimensions. Nevertheless, for the highest mechanical stress value simulated with the 2D continuum model, the experimental resistance trend is accurately reproduced (Figure 4c). Figure 4d presents the filament morphologies for the IRS and LRS obtained from the 2D continuum model, with filaments A, B, and C corresponding to the simulated *I–V* characteristics shown in Figure 4b. All filaments are approximately the same height (4.7 nm). With decreasing lateral mechanical stress, the morphological differences progressively increase, eventually exceeding a factor of 2 in the difference in the filament radius.

**Interpretation and Mechanistic Proposal**

Based on a comparison of the reproducibility of measurement data by both simulation models, we propose a mechanistic framework for filamentary switching in our layered 2D MoS$_2$ ECM memristors.

In Figure 4b, the compact model exhibits no switching during the IRS ↔ LRS transition, which can be explained by its lack of modeling the lateral filament growth. Since the minimum gap is constrained to physically meaningful dimensions > 0.07 nm and under the assumption of single filament formation, no further reduction in resistance is achievable.

In contrast, the 2D model accounts for vertical and lateral filament growth, active electrode dissolution, and, critically, lateral mechanical stress. It successfully reproduces the experimental data, indicating that lateral filament expansion is a key mechanism in layered 2D MoS$_2$ ECM memristors. We propose that desulfurization of MoS$_2$,[57] which is energetically favored in the basic environment generated by water splitting[58] at the electrodes during ECM operation—an essential counter electrode reaction in ECM cells[59–63]—is integral to the switching mechanism. Given the known catalytic activity of MoS$_2$ in water splitting, we suggest that during HRS ↔ IRS cycling, progressive desulfurization is facilitated by the interplay of mechanical stress, electrothermal and tunneling-barrier heating, polarization effects, and water splitting, proceeding through the following redox reactions:

$$S + 2H^+ + 2e^- \rightarrow H_2S \uparrow \tag{3}$$

$$S + 6OH^- \rightarrow SO_3^{2-} + 3H_2O + 4e^- \tag{4}$$

$$MoS_2 \xrightarrow[H_2O, \sigma_{mech}]{E, \Delta T} Mo^{4+} + 2S \uparrow \tag{5}$$

Equations (3) and (4) represent possible S reaction pathways, while equation (5) summarizes the MoS$_2$ decomposition reaction, where $E$ denotes the electric field, $\Delta T$ represents the local temperature increase during switching, $\sigma_{mech}$ represents the mechanical stress induced by filament formation, and H$_2$O encompasses water dissociation at the CF surface, leading to electrochemical decomposition of MoS$_2$.

It is suggested that after extensive HRS ↔ IRS cycling, when the memristor can no longer be reset, the resulting CF consists predominantly of Ag with incorporated Mo, originating from the desulfurization-induced breakdown of the $MoS_2$ lattice. This structural transformation renders the electrochemical dissolution of the filament unfeasible, thereby inhibiting reset functionality while still allowing further filament growth under continued electrical bias.

During the subsequent IRS ↔ LRS switching, we propose that the layered structure of $MoS_2$ plays a critical role by enabling the formation of layer-like dendrites within the vdW gaps of the intact $MoS_2$ layers, where the ionic conductivity is greater due to the anisotropic transport properties of the material. These dendrites do not disrupt the $MoS_2$ lattice structure, allowing it to be reversibly dissolved during the reset process. This results in a composite filament morphology, featuring a bulky core formed during IRS switching, surrounded by laterally extended, resettable dendritic structures confined to the vdW gaps during IRS ↔ LRS switching.

The simulated filament morphologies presented in Figure 4d represent the envelope of the conductive region, opening the possibility for various internal structural configurations.

The schematics in Figures 5a and 5b illustrate the ECM cell structure with layered $MoS_2$ and the proposed internal filament morphologies for the IRS and LRS states, respectively. We suggest a bulky, fully formed core structure for the IRS, whereas dendritic arms and layered features develop within the vdW gaps of the 2D $MoS_2$ for the LRS, as illustrated in Figure 5c. These dendrites can form quantum point contacts with the Ag TE and drastically increase the area of the conductive filament, thereby increasing the current and reducing the resistance. Based on our data and interpretation, we propose that lateral filament growth is particularly pronounced in layered 2D materials such as $MoS_2$, where the vdW gaps act as preferential channels for $Ag^+$ ion migration and filament expansion, giving rise to distinct, abrupt, or gradual RS behavior governed by the intrinsic anisotropic properties of the material.

## Artificial Synapse Applications

We investigated the potential of our Ag/MoS$_2$/Pd vertical memristors for neuromorphic computing applications. Structurally, these devices resemble biological synapses, as illustrated in Figure 6a, where TEs and BEs mimic the presynaptic and postsynaptic terminals, respectively. Functionally, memristors can emulate key synaptic plasticity features, including short-term plasticity (STP) and long-term plasticity (LTP), which are fundamental to learning and memory processes in biological systems.[9,64]

We first investigated the set time ($t_{set}$) for HRS ↔ IRS and IRS ↔ LRS transitions under single-voltage pulses with varying pulse amplitudes ($V_p$). Here, $t_{set}$ is defined as the elapsed time from the onset of the voltage pulse to the point at which the device resistance abruptly decreases to a lower resistive state. Figure S13 illustrates an example of IRS ↔ LRS switching under a pulse with a $V_p$ of 0.4 V. As shown in Figure 6b, increasing the $V_p$ markedly shortens $t_{set}$ from several hundred microseconds to the nanosecond regime, highlighting the ultrafast switching capability and potential for low-power operation.[15,19,28] In the context of STP, we examined two prototypical synaptic phenomena, paired-pulse facilitation (PPF) and spike-rate-dependent plasticity (SRDP), in which prior stimuli enhance subsequent synaptic responses. Note that PPF, SRDP, and the STP-to-LTP transition were studied in a memristor in the HRS. PPF was characterized by applying two consecutive pulses. The PPF index (Figure 6c), defined as the ratio of the second pulse current to the first pulse current (PPF index = $I_2/I_1$), exhibited exponential decay with increasing pulse Δt in the following form: $y = C_1 exp(-x/\tau) + y_0$. This behavior is consistent with the short-term synaptic plasticity observed in biological systems.[65] The highest PPF index was observed at Δ$t$ = 0.2 μs, whereas facilitation became negligible when Δ$t$ exceeded 10 μs. The details are shown in Figure S14.

We then investigated the frequency dependence of the synaptic response (the SRDP behavior) by applying trains of ten pulses. As shown in Figure 6d, the synaptic current response varied markedly with the interstimulus Δ$t$. At a short Δ$t$ of 0.2 μs, strong facilitation was observed without saturation. For intermediate Δ$t$ (0.5 μs and 1 μs), facilitation occurred but rapidly reached saturation. At longer Δ$t$ values (5 μs and 10 μs), facilitation was not evident.

Next, we explored the transition from STP to LTP by varying the number of applied pulses. When 20 consecutive pulses ($V_p$ = 0.4 V, Δ$t$ = 1 μs, pulse width ($t_p$) = 0.5 μs) were applied, the current gradually increased during each pulse and subsequently relaxed, indicative of STP behavior (Figure 6d). However, applying 40 consecutive pulses ($V_p$ = 1 V, Δ$t$ = 0.5 μs, $t_p$ = 1 μs) induced a sharp and persistent increase in current (> 5 mA), indicating a transition to LTP (Figure 6e). This transition can be attributed to the tunable volatile and nonvolatile switching behavior and filament formation dynamics, which are consistent with the Atkinson–Shiffrin model of short- and long-term memory.[66] The conductance values measured during the STP and LTP, as well as after 60 s, are summarized in Figure S15.

We further investigated LTP characteristics in the IRS by analyzing the dynamic current responses to pulses of varying $t_p$ and Δ$t$ (Figure 6g). The current response under 80-pulse stimuli can be categorized into three regimes: (α) subthreshold, where the lateral growth of the filament is inhibited; (β) transition, where lateral growth of the filament initiates; and (γ) saturation, where a filament with laterally grown dendrites is established. A shorter Δ$t$ or a wider $t_p$ significantly reduced the time required to reach the saturation region (γ), demonstrating the effective programmability and plasticity of the memristor.[9]

Notably, the synaptic response time of our memristors is on the order of microseconds, substantially faster than the millisecond-scale of Ca$^+$ ion transmission in biological synapses.[67] This response suggests potential for high-speed neuromorphic computing and memory storage applications.

## Conclusion

In summary, we demonstrated wafer-scale Ag/MoS$_2$/Pd memristors that exhibited both threshold and non-volatile switching behaviors, with low switching voltages in DC measurements and fast $t_{set}$ under PVS. We consistently observed intermediate resistive states in our memristors across the wafer. Structural and compositional analyses, including AFM, TEM, and EDXS mapping, revealed details about the quality and chemical composition of the MOCVD-grown MoS$_2$ films and devices. The devices demonstrated reliable and reproducible IRS ↔ LRS switching with endurance exceeding 2500 switching cycles in DC mode, state retention for at least 10$^6$ s, and low switching variability. Combining EDXS mapping and area-dependent and temperature-dependent measurements revealed a filamentary switching mechanism governed by Ag$^+$ ion migration. Based on an analytical 1D compact model and 2D axisymmetric continuum simulations, we attribute the IRS ↔ LRS transitions to the lateral extension of Ag filaments along vdW gaps following the formation of a vertical remnant filament. This suggests a previously unexplored ionic migration pathway, in which Ag$^+$ ions diffuse laterally within the 2D interlayer gaps and ultimately drive the RS process. Moreover, we experimentally demonstrated synaptic plasticity under pulsed stimuli, highlighting the potential of these devices for neuromorphic computing. Together, these results provide new insights into the RS mechanisms in 2D material-based memristors, while offering a scalable platform for exploring next-generation memory and neuromorphic systems.

# Experimental Section

*Device Fabrication*: The fabrication of Ag/MoS$_2$/Pd memristors includes 3 steps: defining bottom electrodes (BEs), patterning MoS$_2$, and defining top electrodes (TEs). The memristors were fabricated on a 6-inch wafer with 90 nm thermal silicon dioxide (SiO$_2$). Initially, BEs were defined via stepper lithography (Canon i-line Stepper) with an AZ NLOF 2020 Photoresist (MicroChemicals GmbH). 5 nm titanium (Ti) (as the adhesion layer) and 45 nm palladium (Pd) were deposited via electron-beam evaporation in an FHR evaporation tool (FHR. Star. 200-EVA) following a lift-off process in dimethyl sulfoxide (DMSO) at $T$ = 80 °C for 10 min. The multilayer MoS$_2$ material used for this work was grown via metal–organic chemical vapor deposition (MOCVD) on 2-inch sapphire developed by Grundmann et al.[68] The material was transferred onto the patterned BEs via a wet transfer method with poly(methyl methacrylate) (PMMA) as a supporting layer. The PMMA/MoS$_2$ layer was delaminated from the as-grown sapphire wafer via a 4 mol potassium hydroxide (KOH) solution and cleaned from the etchant by floating on DI water overnight. After transfer, the PMMA on the surface was removed in acetone at $T$ = 80 °C for 30 min. Optical lithography was then used to pattern the MoS$_2$ with an AZ MIR 701 Photoresist (MicroChemicals GmbH). Subsequent patterning of MoS$_2$ was achieved through reactive ion etching (RIE) in an Oxford Instruments "Plasmalab System 100" tool with a mixture of tetrafluoromethane (CF$_4$) and oxygen gas. After patterning MoS$_2$ on the BEs, stepper optical lithography was applied again using an AZ NLOF 2020 Photoresist to define the TEs. Subsequently, the 50 nm Ag was evaporated, and a lift-off process was performed in DMSO at room temperature.

*Material and Device Characterization*: Optical microscope images were recorded with a Leica INM100 microscope. Atomic force microscopy (AFM) was performed on the as-grown MOCVD MoS$_2$ from the sapphire wafer with a Dimension Icon AFM from Bruker Instruments using a tip with 26 N/m as the force constant and a frequency variation between 200 and 400 kHz (OTESPA) in tapping mode in air and at room temperature. Raman measurements were conducted via an alpha300R WITec confocal Raman spectrometer with an excitation laser wavelength of 532 nm and a laser power of 1 mW. The measurements were performed in mapping mode at room temperature. The Raman resolution was 1800 g·mm$^{-1}$ grating. Transmission electron microscopy (TEM) was performed with an FEI Titan G2 80-200 ChemiSTEM microscope with an accelerating voltage of 200 kV. The microscope is equipped with an extreme-brightness cold field emission gun (XFEG), a probe Cs corrector, a super-X EDXS system, and a Gatan Enfinium ER (model 977) spectrometer. Focused ion beam (FIB) lamellas were prepared from the fabricated memristors via an FEI Helios Nanolab 660 dual-beam microscope with gallium (Ga) ions and fixed on a copper (Cu) FIB lift-out grid. The energy-dispersive X-ray spectroscopy (EDXS) chemical mapping was approximately 22 mrad, whereas the collection semiangles were 80-200 mrad for high-angle annular dark-field (HAADF) imaging. EDXS maps were typically collected for approximately 30 minutes, and background subtraction was performed. The composition of the as-grown MOCVD MoS$_2$ was analyzed via X-ray photoelectron spectroscopy (XPS) with a Thermo Scientific KA1066 spectrometer and monochromatic Al K-α X-rays at 1486.6 eV. Charge correction was applied to position the C–C bonding contribution from adventitious carbon at 284.8 eV to correct for potential charging effects in the XPS spectra. Contributions from S 2p and Mo 3d were measured, and the atomic S to Mo ratios were calculated from the integrated peak areas via the appropriate sensitivity factors.

*Electrical Measurements*: Electrical measurements were conducted in a cryogenic probe station (LakeShore) connected to a semiconductor parameter analyzer (SPA) 4200A-SCS by Tektronix. Current-voltage (*I-V*) measurements were performed in DC mode at room temperature (21 °C) using two source measurement unit (SMU) cards (Keithley 4200-SMU), each connected to a preamplifier (Keithley 4200-PA). The internal current compliance of the semiconductor parameter analyzer was used to limit the current during the SET transition to avoid permanent breakdown. During the SET transition, the voltage automatically decreases as soon as the current reaches the current compliance ($I_{cc}$). This leads to a part of the curve where the voltage decreases while the current increases, which should not be misinterpreted as a negative differential resistance. The $I_{cc}$ was deactivated for the reset sweeps to reach the currents required to induce the transitions. Temperature-dependent measurements were performed in vacuum ($8 \times 10^{-5}$ mBar), and a forward sweep from -0.2 V to +0.2 V was applied. Pulse measurements for artificial synaptic applications were conducted via pulse measurement units (Keithley 4225-PMUs), where each channel was connected to a remote amplifier unit "Keithley 4225-RPM". Throughout this work, the rise and fall times for each individual pulse were consistently set to 20 ns, and the current response was extracted at the end of a pulse.

# AUTHOR INFORMATION

Corresponding Author

Max C. Lemme − AMO GmbH, 52074 Aachen, Germany; Chair of Electronic Devices, RWTH Aachen University, 52074 Aachen, Germany; orcid.org/0000-0003-4552-2411; Email: max.lemme@rwth-aachen.de

Authors

Yuan Fa − AMO GmbH, Otto-Blumenthal-Str. 25, 52074 Aachen, Germany; Chair of Electronic Devices, RWTH Aachen University, 52074 Aachen, Germany; orcid.org/0009-0008-7143-9236

Milan Buttberg − Institute of Materials in Electrical Engineering and Information Technology II (IWE2), RWTH Aachen University, Sommerfeldstraße 24, 52074 Aachen, Germany; orcid.org/0000-0003-1647-4254

Ke Ran − AMO GmbH, Otto-Blumenthal-Str. 25, 52074 Aachen, Germany; Central Facility for Electron Microscopy, RWTH Aachen University, Ahornstr. 55, 52074 Aachen, Germany; Ernst Ruska-Centre for Microscopy and Spectroscopy with Electrons (ER-C 2), Forschungszentrum Jülich GmbH, Wilhelm-Johnen-Str., 52425 Jülich, Germany; orcid.org/0000-0002-9762-4586

Rana Walied Ahmad − Peter Grünberg Institute 7 (PGI 7) and JARA-FIT, Forschungszentrum Jülich GmbH, Wilhelm-Johnen-Straße, 52428 Jülich, Germany; orcid.org/0000-0002-0445-8110

Dennis Braun − Chair of Electronic Devices, RWTH Aachen University, Otto-Blumenthal-Str. 25, 52074 Aachen, Germany; orcid.org/0000-0003-2803-1784

Lukas Voelkel − Chair of Electronic Devices, RWTH Aachen University, Otto-Blumenthal-Str. 25, 52074 Aachen, Germany; orcid.org/0000-0002-8138-9980

Jimin Lee − Chair of Electronic Devices, RWTH Aachen University, Otto-Blumenthal-Str. 25, 52074 Aachen, Germany; orcid.org/0000-0002-5877-6958

Sofía Cruces − Chair of Electronic Devices, RWTH Aachen University, Otto-Blumenthal-Str. 25, 52074 Aachen, Germany; orcid.org/0000-0003-2816-7016

Bart Macco − Department of Applied Physics and Science Education, Eindhoven University of Technology, 5600 MB Eindhoven, The Netherlands; orcid.org/0000-0003-1197-441X

Bárbara Canto − AMO GmbH, Otto-Blumenthal-Str. 25, 52074 Aachen, Germany; orcid.org/0000-0001-5885-9852

Holger Lerch − AMO GmbH, Otto-Blumenthal-Str. 25, 52074 Aachen, Germany

Thorsten Wahlbrink − AMO GmbH, Otto-Blumenthal-Str. 25, 52074 Aachen, Germany, orcid.org/0000-0001-8963-9828

Holger Kalisch − Compound Semiconductor Technology, RWTH Aachen University, 52074 Aachen, Germany

Michael Heuken − Compound Semiconductor Technology, RWTH Aachen University, 52074 Aachen, Germany; AIXTRON SE, 52134 Herzogenrath, Germany; orcid.org/0000-0001-9739-9692

Andrei Vescan − Compound Semiconductor Technology, RWTH Aachen University, 52074 Aachen, Germany; orcid.org/0000-0001-9465-2621

Joachim Mayer – Central Facility for Electron Microscopy, RWTH Aachen University, Ahornstr. 55, 52074 Aachen, Germany; Ernst Ruska-Centre for Microscopy and Spectroscopy with Electrons (ER-C 2), Forschungszentrum Jülich GmbH, Wilhelm-Johnen-Str., 52425 Jülich, Germany; orcid.org/0000-0003-3292-5342

Zhenxing Wang – AMO GmbH, Otto-Blumenthal-Str. 25, 52074 Aachen, Germany; orcid.org/0000-0002-2103-7692

Ilia Valov – Peter Grünberg Institute 7 (PGI 7) and JARA-FIT, Forschungszentrum Jülich GmbH, Wilhelm-Johnen-Straße, 52428 Jülich, Germany; "Acad. Evgeni Budevski" IEE-BAS, Bulgarian Academy of Sciences (BAS), Acad. G. Bonchev Str, Block 10, Sofia 1113, Bulgaria; orcid.org/0000-0002-0728-7214

Stephan Menzel – Peter Grünberg Institute 7 (PGI 7) and JARA-FIT, Forschungszentrum Jülich GmbH, Wilhelm-Johnen-Straße, 52428 Jülich, Germany; orcid.org/0000-0002-4258-2673

Complete contact information is available at:

**Notes**

The authors declare no competing financial interest.

# ACKNOWLEDGEMENT

This work has received funding from the German Federal Ministry of Research, Technology and Space (BMFTR) within the projects NeuroSys (No. 03ZU1106AA, 03ZU1106AB, 03ZU1106AD, 03ZU2106AA, 03ZU2106AB, 03ZU2106AD, and 03ZU2106AE) and NEUROTEC 2 (No. 16ME0398K, 16ME0399, 16ME0400, and 16ME0403). We acknowledge funding from the European Union's

Horizon Europe research and innovation program (via CHIPS-JU) under the project ENERGIZE (101194458). We acknowledge that this study was supported in part by the Deutsche Forschungsgemeinschaft (DFG) under Project SPP2262 MEMMEA, Project No. 441918103 (MemrisTec).

## ASSOCIATED CONTENT

SI Supporting Information

The Supporting Information is available free of charge at

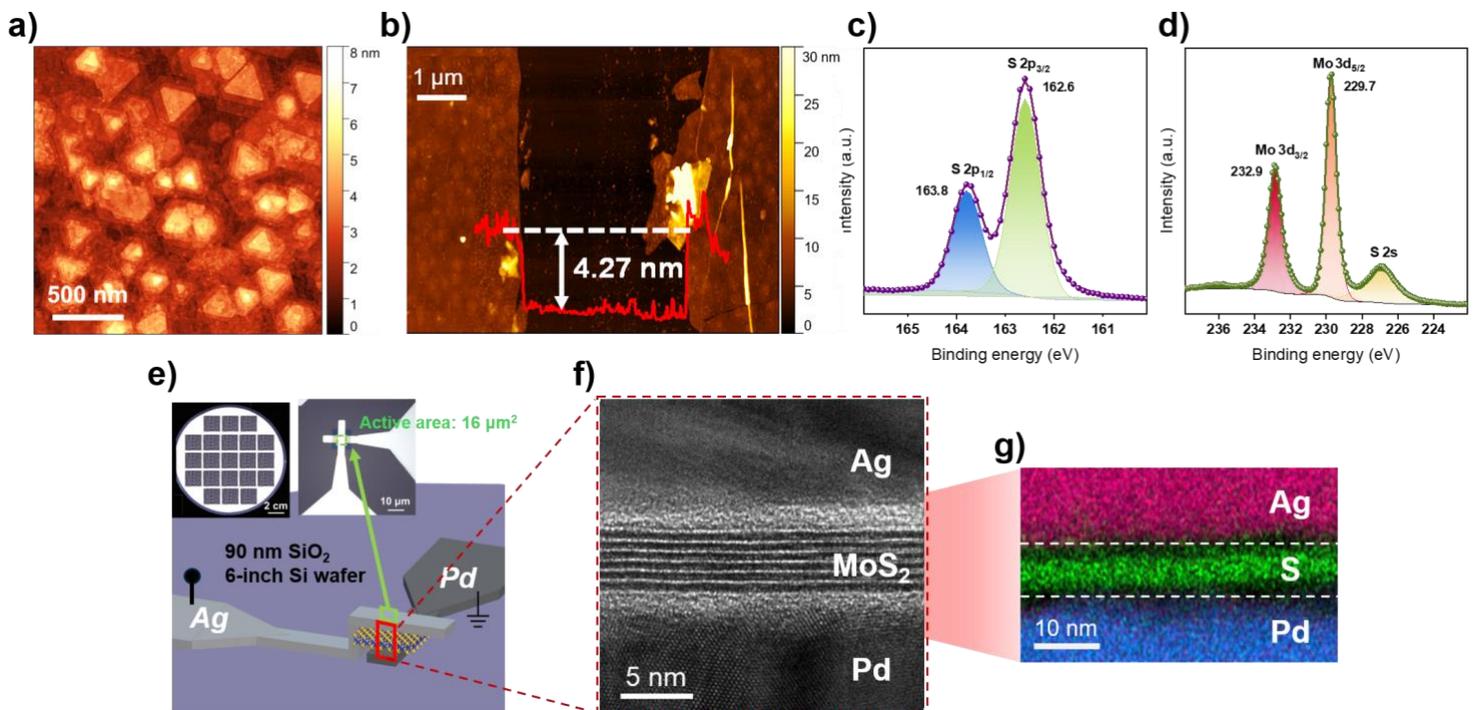

Figure 1: Material analysis. AFM images of the as-grown MOCVD MoS$_2$ (a) surface and (b) edge on a 2-inch sapphire wafer. The inset height profile indicates a thickness of ~4.27 nm. XPS spectra of (c) S 2p and (d) Mo 3d. The S-to-Mo atomic ratio is ~2, which is calculated from the integrated peak areas of the Mo 3d and S 2p peaks. (e) Schematic of a cross-point Ag/MoS$_2$/Pd memristor with an active area of 4 × 4 µm² on a 6-inch Si wafer with 90 nm SiO$_2$ dielectrics. The inset optical images show the fabricated 6-inch wafer (left) and a single cross-point device (right). (f) HRTEM cross-sectional image of a memristor before electrical measurement and (g) EDX elemental maps of Ag, S, and Pd of the same region.

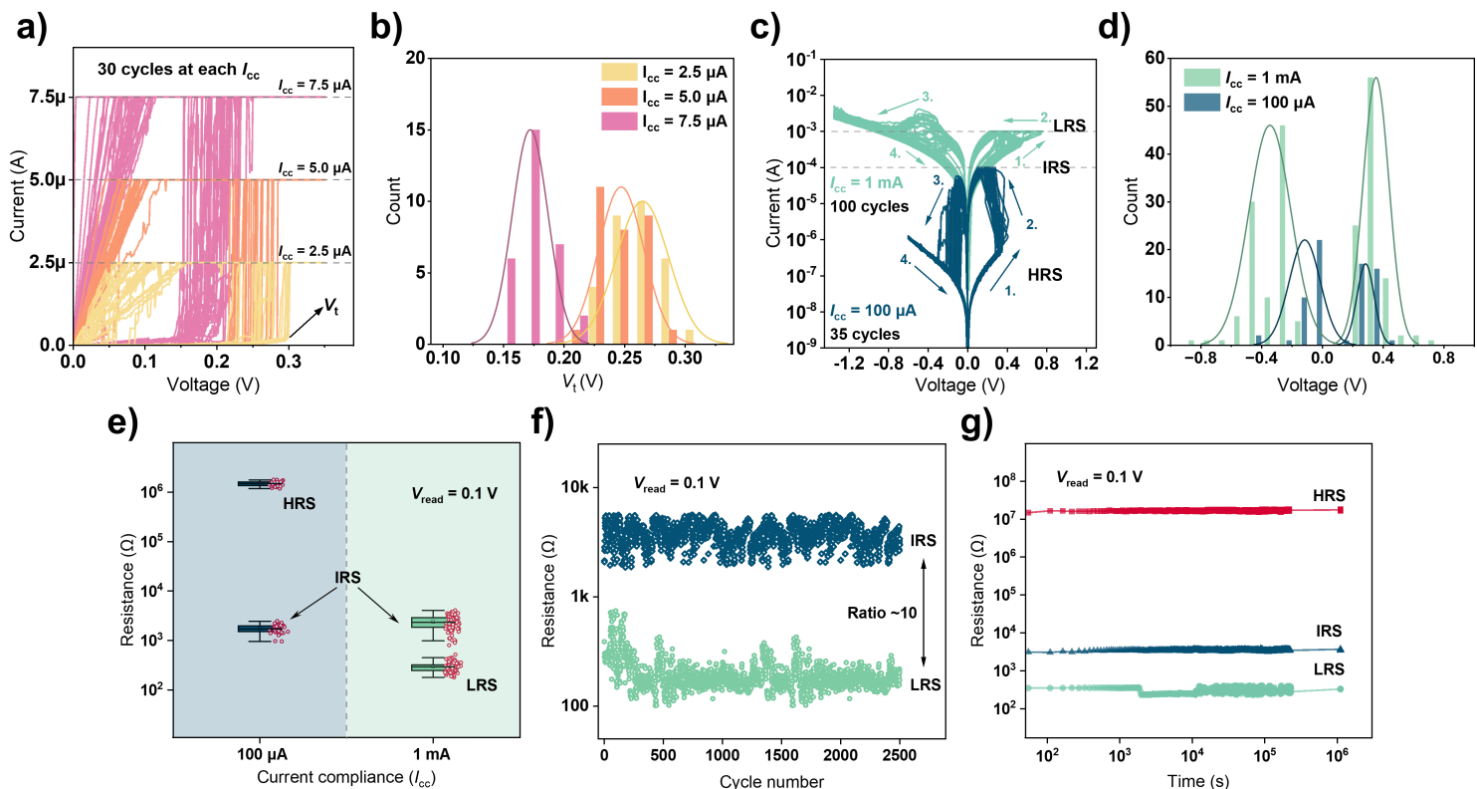

Figure 2: Electrical measurements of memristors. (a) 30 *I–V* sweeps on the memristor under current limits of 2.5 μA, 5.0 μA, and 7.5 μA, respectively, showing TS. The gradually decreasing $V_t$ may be attributed to the measurement sequence. (b) Histogram plot of $V_t$ extracted from (a), fitted with a Gaussian distribution. (c) NVS characteristic curves of the device. Under an $I_{cc}$ of 100 μA (dark curves), the device shows 35 NVS curves with an abrupt set process (HRS ↔ IRS), after which switching ceases at this $I_{cc}$. Upon increasing $I_{cc}$ to 1 mA (light curves), the device exhibited at least 100 stable NVS cycles (IRS ↔ LRS). (d) Histogram plot of $V_{set}$ and $V_{reset}$ extracted from (c), fitted with a Gaussian distribution. (e) Extracted resistances of the HRS, IRS, and LRS under $I_{cc}$ of 100 μA and 1 mA from (c). (f) Endurance performance of a memristor showing 2500 cycles of *I–V* sweeps in DC mode. (g) State retention measurement showing that each resistive state remains stable for more than $10^6$ s.

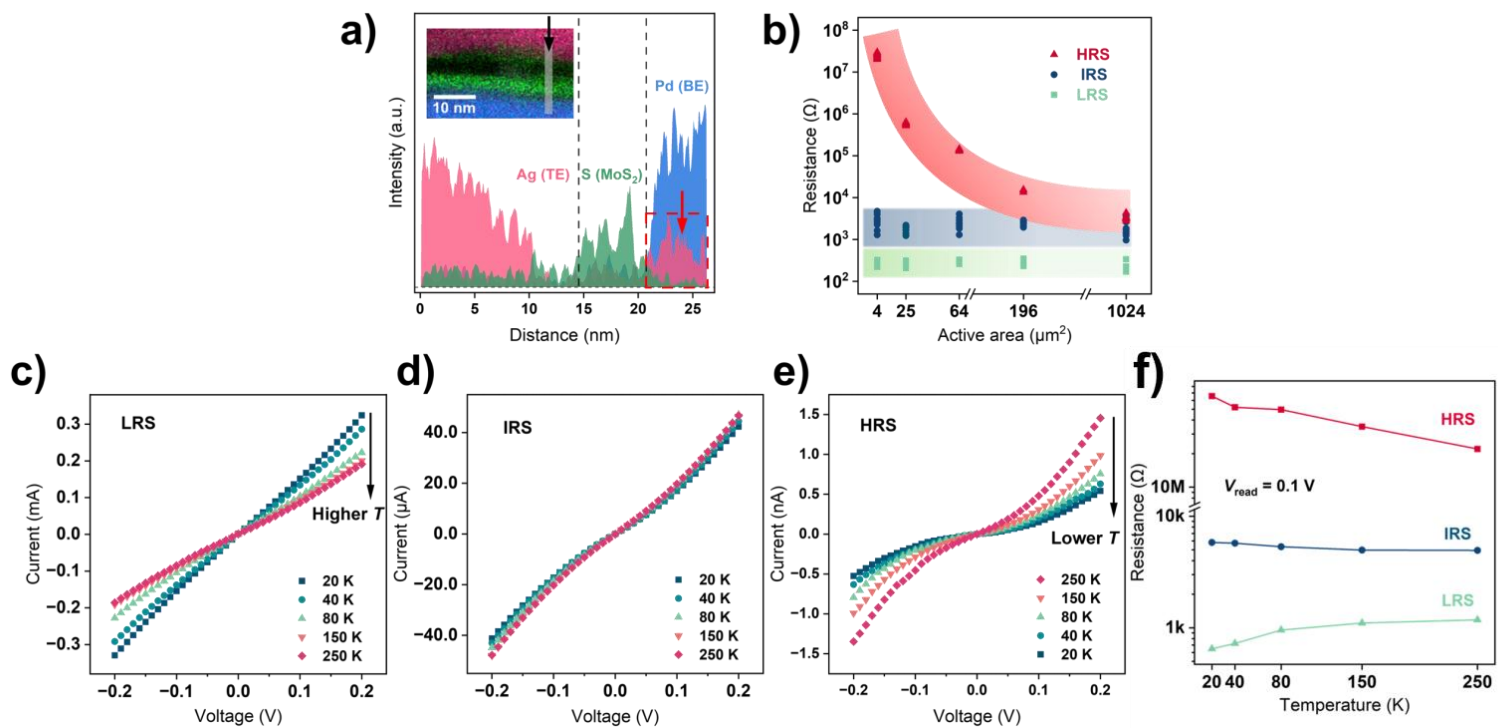

Figure 3: Conduction mechanisms. (a) Elemental line scan across the cross-section of a memristor after electrical measurements. The pronounced Ag signal in the BE region indicates the migration of Ag$^+$ ions. The valley of the S peak aligns with the maximum of the Ag peak (marked with the back dashed lines), indicating lateral diffusion of Ag within the vdW gaps between MoS$_2$ layers. The inset EDXS map shows the elemental distribution and the scanned region. (b) Resistance versus active area plot of five devices (4–1024 µm²). The area-dependent HRS and area-independent IRS and LRS indicate a filamentary switching mechanism. I–V curves of the devices in the (c) LRS, (d) IRS, and (e) HRS measured at temperatures ranging from 20 K to 250 K. (e) Resistance versus temperature plot at a $V_{read}$ of 0.1 V, showing distinct temperature dependencies for different resistive states.

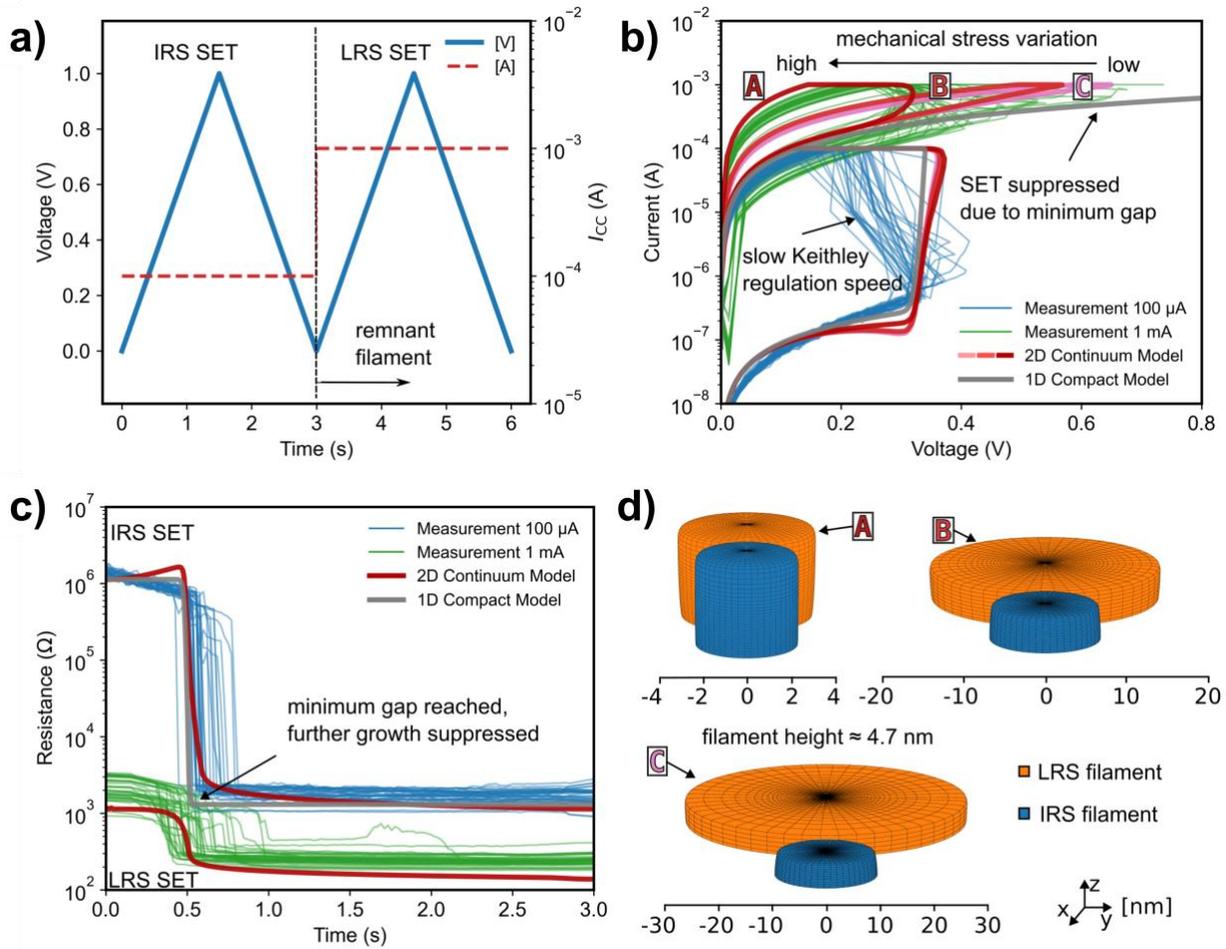

Figure 4: Simulation and modeling of IRS–LRS switching behavior. (a) Applied bias scheme: two consecutive triangular sweeps with $I_{CC}$ of 100 μA and 1 mA, assuming a remnant filament after the first sweep. (b) Simulated and measured *I–V* characteristics: comparison between experimental data, the 1D compact model (gray), and the 2D continuum model (A–C). The compact model fails to reproduce IRS ↔ LRS transitions due to premature gap closure, while the continuum model captures variability via lateral mechanical stress effects. (c) Resistance evolution from experiment and simulations. (d) Filament morphologies from the 2D continuum model for cases A–C, corresponding to decreasing lateral mechanical stress (A: high, C: low). Labels A–C in (d) correspond to the I–V curves in (b).

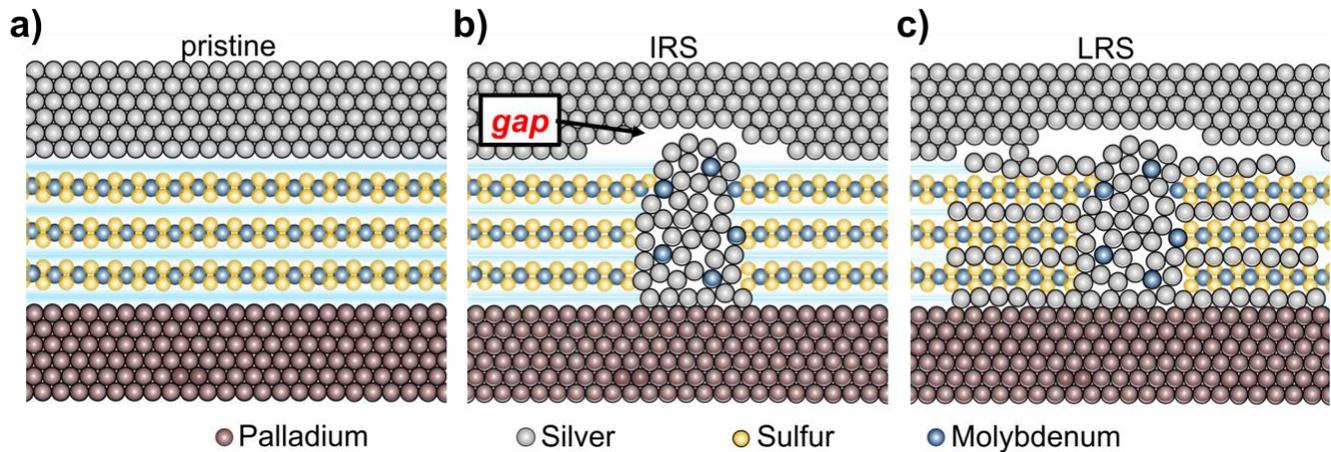

Figure 5: Schematic of the proposed filament formation mechanisms. Three different internal states of the reported 2D $MoS_2$ cell: (a) pristine, (b) IRS, and (c) LRS. During IRS switching, the $MoS_2$ lattice is disrupted due to desulfurization, inhibiting further filament dissolution. During IRS ↔ LRS switching, the mechanism changes due to the anisotropic ionic conductivity of the 2D material. Enhanced ionic transport along the lateral vdW gaps promotes the formation of dendrites, resulting in a composite filament consisting of a mixed Ag/Mo bulk material with Ag dendrites extending laterally to further reduce the resistance. For clarity, only three $MoS_2$ layers are shown in the schematic. The gap between the active electrode and the filament is indicated in (b).

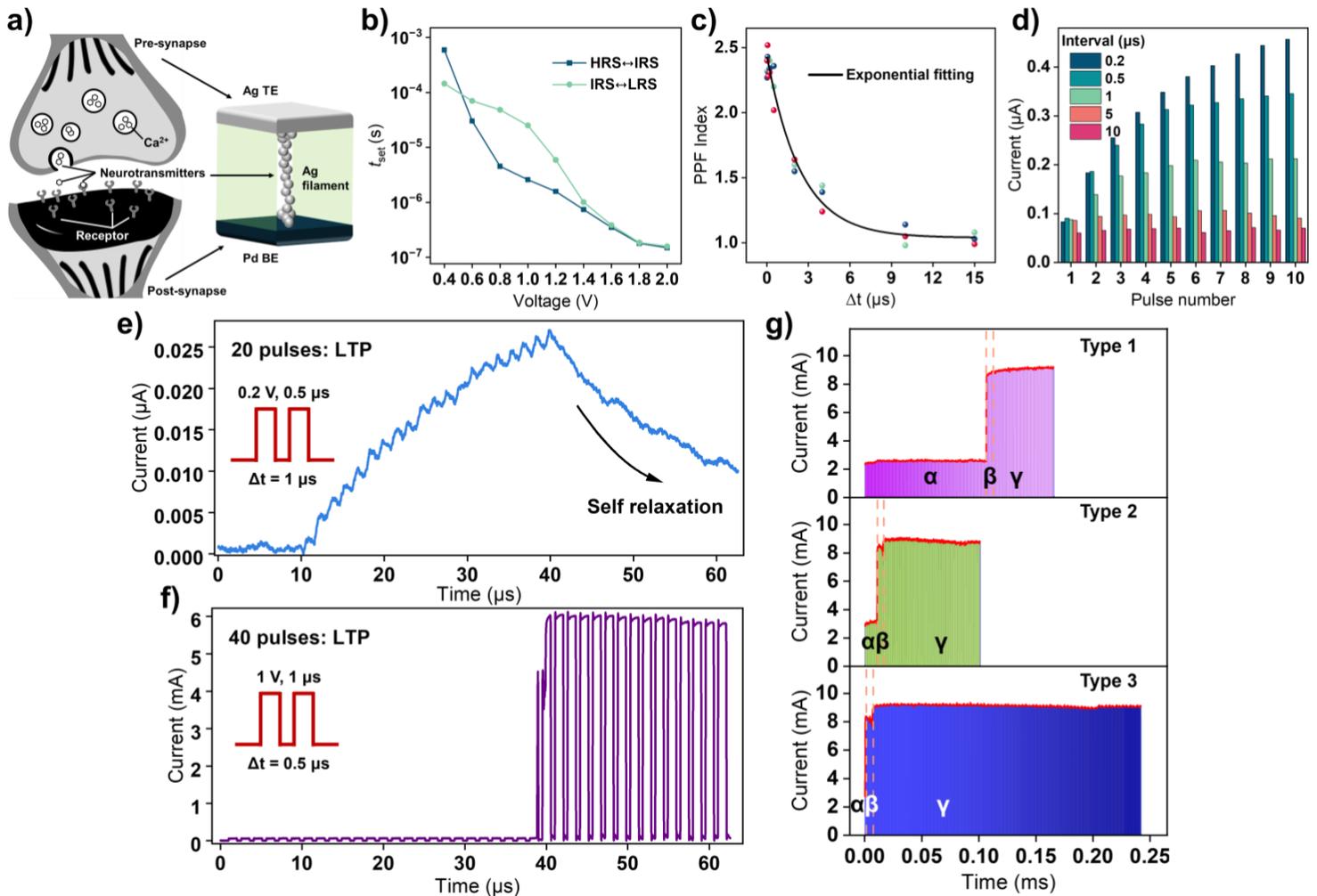

Figure 6: Artificial synapse demonstration. (a) Schematics of a biological synapse and a memristor, illustrating their structural and functional similarities. (b) $t_{set}$ for HRS ↔ IRS and IRS ↔ LRS switching under single-voltage pulses with $V_p$ ranging from 0.4 to 2 V. Increasing $V_p$ reduces $t_{set}$ to the nanosecond regime, highlighting the potential for low-power operation. (c) $\Delta t$-dependent PPF behavior of the device measured by two consecutive pulses. (d) $\Delta t$-dependent SRDP behavior measured by ten consecutive pulses. (e) Current response under 20 consecutive pulses followed by a current self-reaction, showing STP. (f) Current response under 40 consecutive pulses with a sudden increase in current level, showing LTP. (g) The dynamic response of the device in the IRS, where a larger $t_p$ and shorter $\Delta t$ accelerate the approach to the saturation current level.

**TOC**

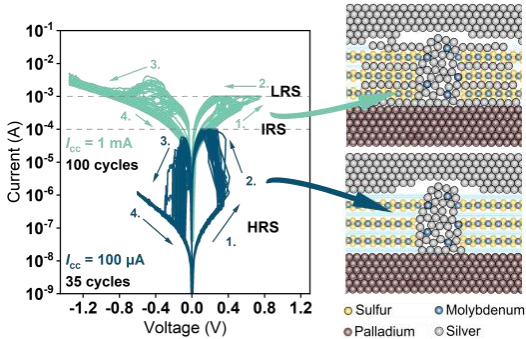

# Supporting Information

# Intermediate Resistive State in Wafer-Scale MoS$_2$ Memristors through Lateral Silver Filament Growth for Artificial Synapse Applications

Yuan Fa[1,2], Milan Buttberg[3], Ke Ran[1,4,5], Rana Walied Ahmad[6], Dennis Braun[2], Lukas Völkel[2], Jimin Lee[2], Sofía Cruces[2], Bart Macco[7], Bárbara Canto[1], Holger Lerch[1], Thorsten Wahlbrink[1], Holger Kalisch[8], Michael Heuken[8,9], Andrei Vescan[8], Joachim Mayer[4,5], Zhenxing Wang[1], Ilia Valov[6,10], Stephan Menzel[6], and Max C. Lemme[1,2*]

1 AMO GmbH, Advanced Microelectronic Center Aachen, Otto-Blumenthal-Str. 25, 52074 Aachen, Germany

2 Chair of Electronic Devices (ELD), RWTH Aachen University, Otto-Blumenthal-Str. 25, 52074 Aachen, Germany

3 Institute of Materials in Electrical Engineering and Information Technology II (IWE2), RWTH Aachen University, Sommerfeldstraße 24, 52074 Aachen, Germany

4 Central Facility for Electron Microscopy (GFE), RWTH Aachen University, Ahornstr. 55, 52074 Aachen, Germany

5 Ernst Ruska-Centre for Microscopy and Spectroscopy with Electrons (ER-C 2), Forschungszentrum Jülich GmbH, Wilhelm-Johnen-Str., 52425 Jülich, Germany.

6 Peter Grünberg Institute 7 (PGI 7) and JARA-FIT, Forschungszentrum Jülich GmbH, Wilhelm-Johnen-Straße, 52428 Jülich, Germany

7 Department of Applied Physics and Science Education, Eindhoven University of Technology, 5600 MB Eindhoven, The Netherlands

8 Compound Semiconductor Technology (CST), RWTH Aachen University, Sommerfeldstr. 18, 52074 Aachen, Germany

9 AIXTRON SE, Dornkaulstr. 2, 52134 Herzogenrath, Germany

10 "Acad. Evgeni Budevski" IEE-BAS, Bulgarian Academy of Sciences (BAS), Acad. G. Bonchev Str, Block 10, Sofia 1113, Bulgaria

Email: max.lemme@rwth-aachen.de

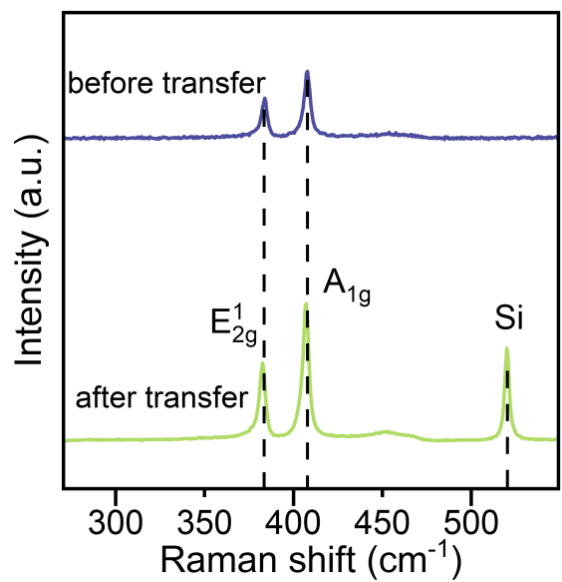

Figure S1: Raman spectra of the MOCVD MoS$_2$ before and after wet transfer. The unchanged peak positions indicate that wet transfer causes no quality degradation to MoS$_2$.

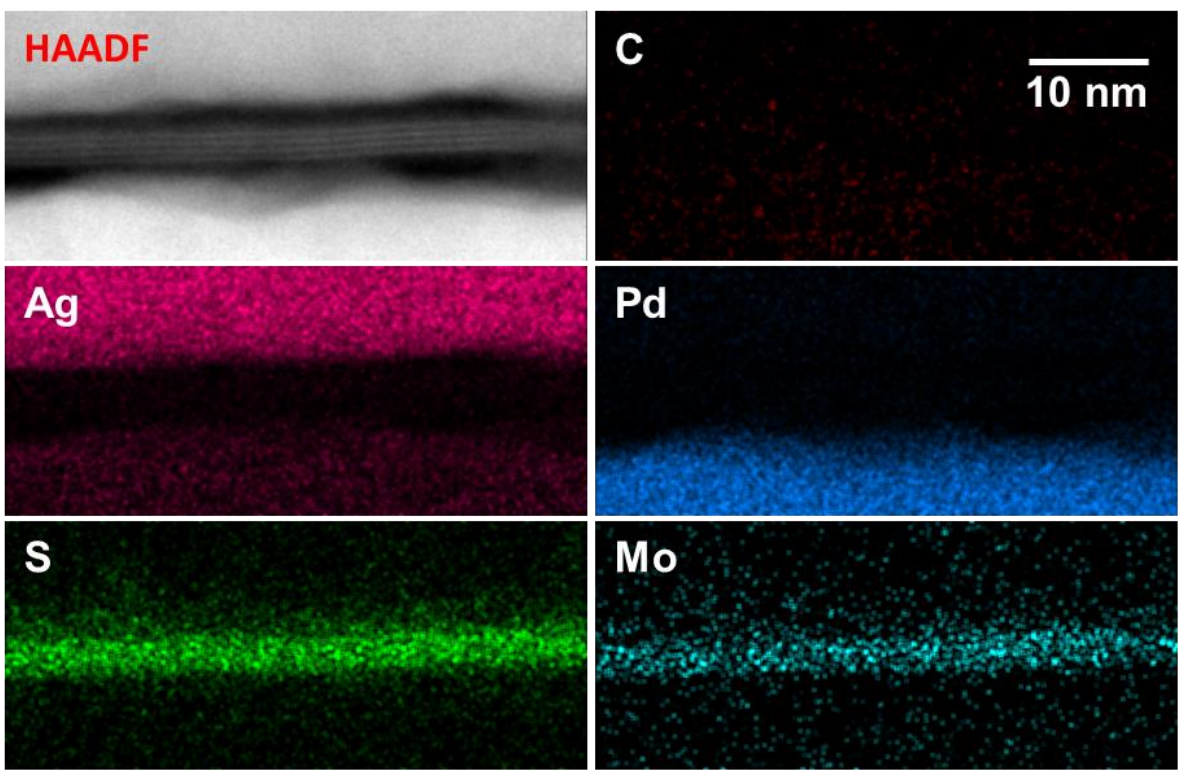

Figure S2: HAADF image and corresponding EDXS elemental maps of C, Ag, Pd, S, and Mo.

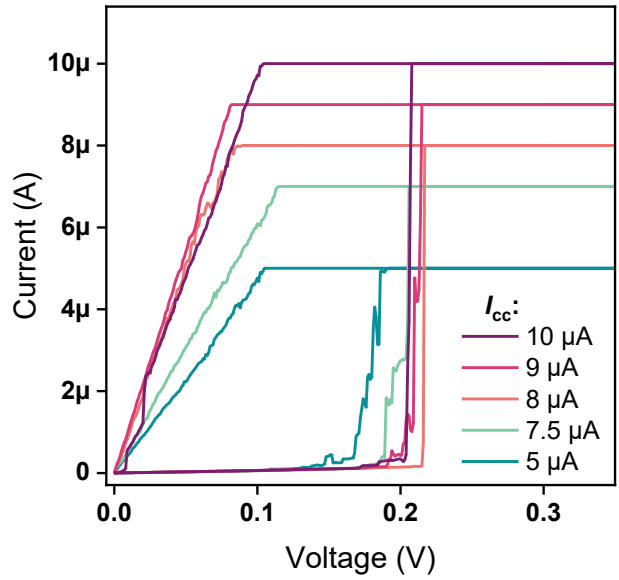

Figure S3： Threshold resistive switching under various $I_{cc}$ values, showing that $V_t$ is independent of $I_{cc}$

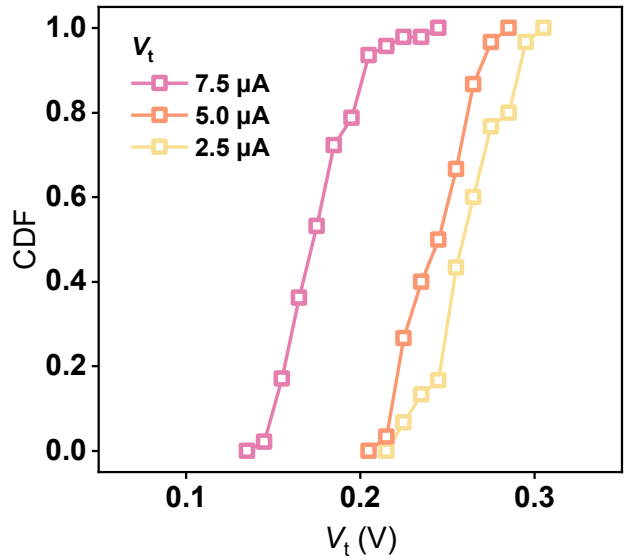

Figure S4: Cumulative distribution functions (CDF) of $V_t$ under $I_{cc}$ of 2.5, 5.0, and 7.5 µA resulting in threshold switching.

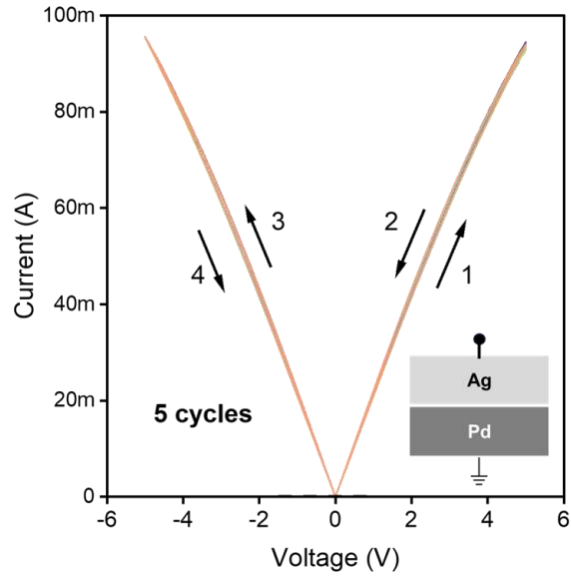

Figure S5: Five consecutive *I–V* curves of a memristor without MoS$_2$, fabricated in the same process as the MoS$_2$-based devices.

Table S1: Statistics of the mean values and standard deviations of $V_t$ during the TS cycles in Figure 2a.

| $I_{cc}$ | Mean value | Standard Deviation |
|---|---|---|
| 2.5 µA | 0.26 | 0.02 |
| 5 µA | 0.25 | 0.02 |
| 7.5 µA | 0.17 | 0.01 |

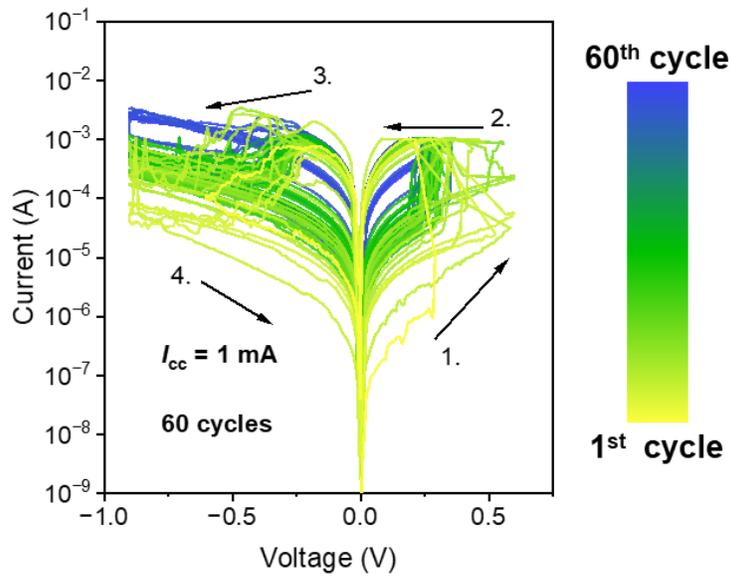

Figure S6: 60 NVS cycles measured with an initial $I_{cc}$ of 1 mA after TS on the same device, exhibiting chaotic switching behavior and a progressively shrinking resistive switching window.

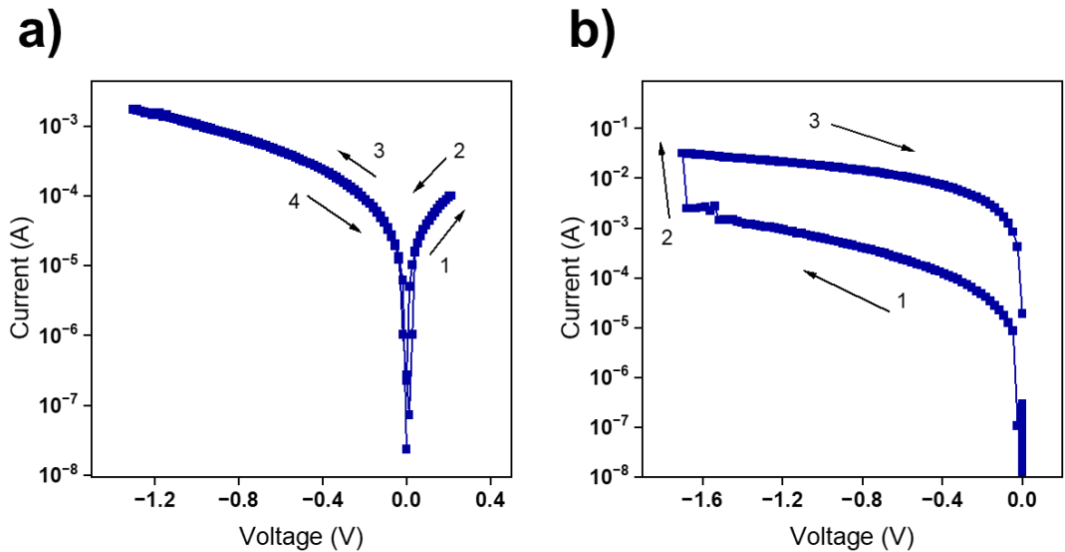

Figure S7: a) *I–V* curve measured after 35 NVS cycles under an $I_{cc}$ of 100 uA, showing that the NVS cannot proceed under such $I_{cc}$ and that the device remains in a permanent IRS. b) *I–V* sweep with a larger negative voltage, leading to permanent shorting of the device.

Table S2: Statistics of the mean values and standard deviations of $V_{set}$ and $V_{reset}$ under $I_{cc}$ of 100 µA and 1 mA in NVS cycles in Figure 2c.

|  | $I_{cc}$ | Mean value | Standard Deviation |
|---|---|---|---|
| $V_{set}$ | 100 µA | 0.29 | 0.06 |
|  | 1 mA | 0.35 | 0.09 |
| $V_{reset}$ | 100 µA | -0.11 | 0.10 |
|  | 1 mA | -0.34 | 0.13 |

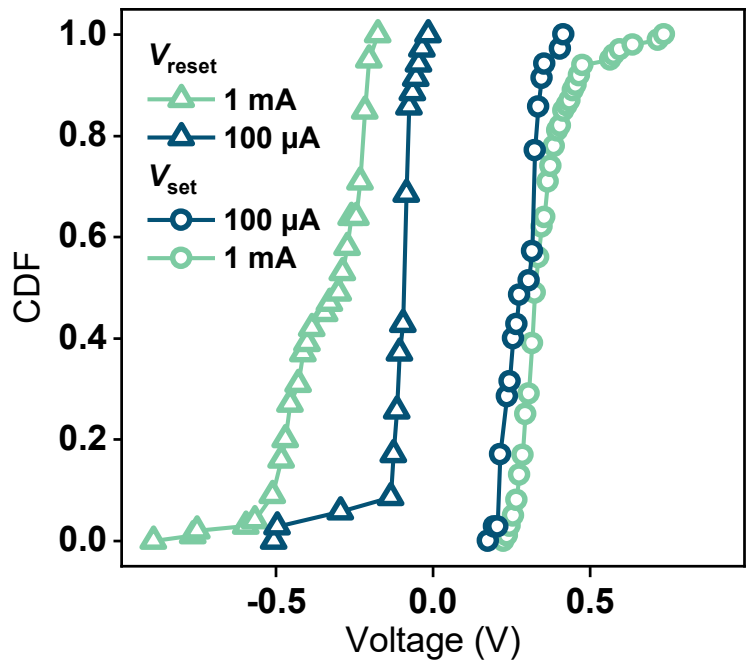

Figure S8: CDFs of $V_{set}$ and $V_{reset}$ under $I_{cc}$ of 100 µA and 1 mA in the NVS.

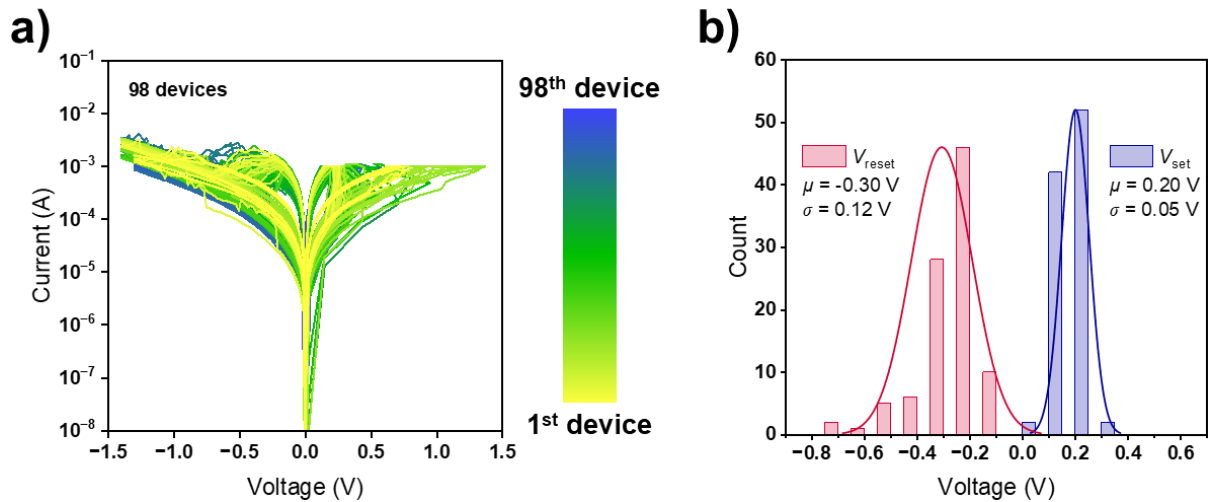

Figure S9: a) First-cycle *I–V* curves showing RS in the IRS ↔ HRS from 98 devices across the wafer. b) Gaussian distributions of the corresponding $V_{set}$ and $V_{reset}$ values extracted from (a).

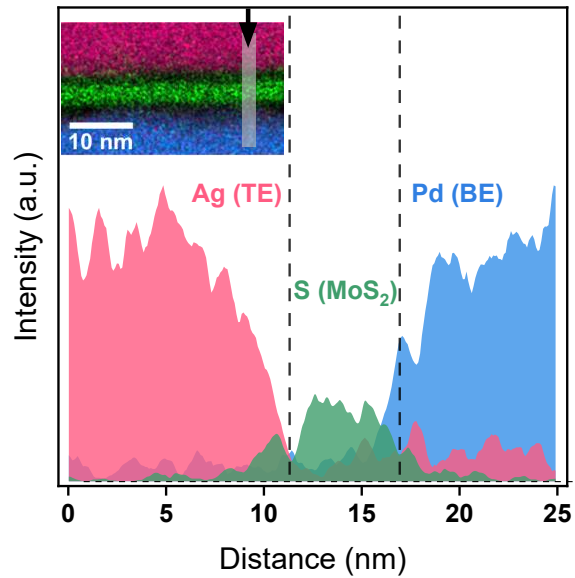

Figure S10: Line scan intensities of the Ag, S, and Pd distributions on the cross-section of the memristor in the HRS (no electrical measurements), showing no $Ag^+$ ion migration

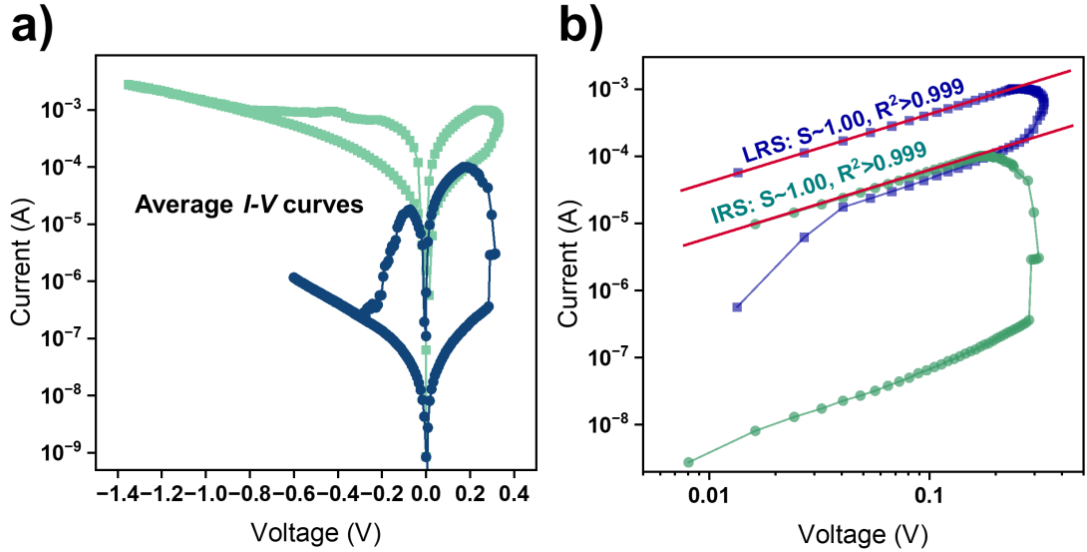

Figure S11: (a) Averaged *I–V* curves extracted from Figure 2c. The darker curve represents the average over 35 cycles with an $I_{cc}$ of 100 µA, whereas the lighter curve corresponds to the average over 100 cycles with an $I_{cc}$ of 1 mA. (b) Double-logarithmic plot of the IRS and LRS with linear fitting. Both states exhibit slopes of ~1 with $R^2 > 0.999$.

Table S3: Extracted slopes and $R^2$ from linear fits of *I–V* sweeps for the memristor in the LRS at various temperatures.

| T (K) | Slope | $R^2$ |
|---|---|---|
| 20 | 0.0016 ± 6.37E-6 | 0.99937 |
| 40 | 0.0014 ± 5.29E-6 | 0.99944 |
| 80 | 0.0011 ± 4.93E-6 | 0.99920 |
| 150 | 9.50E-4 ± 3.96E-6 | 0.99932 |
| 250 | 9.01E-4 ± 5.78E-6 | 0.99904 |

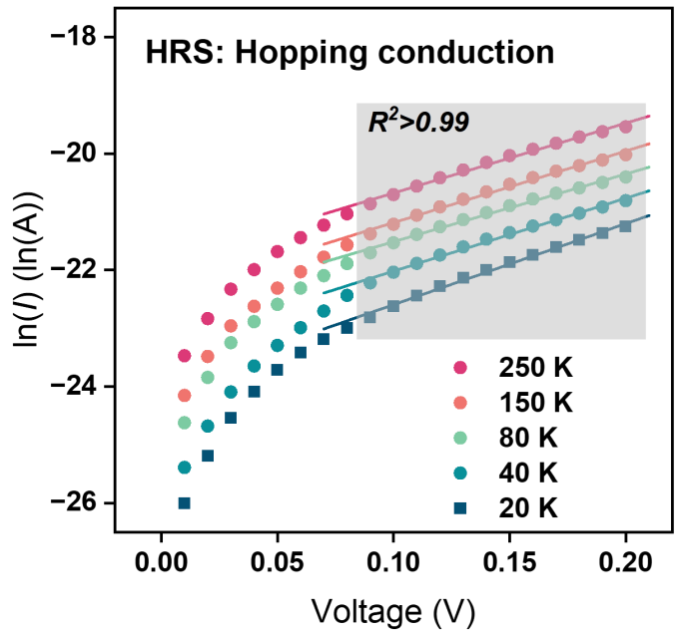

Figure S12: Hopping conduction plot for the memristor in the HRS, corresponding to the data in Figure 3e. Data points in the 0.09-0.2 V range (highlighted with a gray background) exhibit a linear fit $R^2$ exceeding 0.99. A small offset was applied to enhance visibility.

**Switching Mechanism in the HRS**

As the resistances in the HRS clearly depend on temperature (Figure 3d), we infer that hopping conduction may serve as a dominant transport mechanism. In MoS$_2$ films, hopping conduction arises from the tunneling of trapped electrons between two trap sites, effectively reducing the barrier width. The current density $J$ in this regime can be described by the following expression:[1]

$$J = qanv\exp\left(\frac{qaE}{k_BT} - \frac{E_a}{k_BT}\right) \quad (1)$$

where $q$ is the elementary charge, a is the average distance between the trap sites, $n$ is the electron concentration in the conduction band, $v$ is the frequency of thermal vibrations of electrons at trap sites, $E$ is the applied electric field, $k_B$ is the Boltzmann constant, $T$ is the temperature, a is the mean distance between the trap sites and $E_a$ is the energy level difference between the trap states and the conduction band minimum. According to Equation (1), a linear relationship is expected between ln($J$) and $E$, which is characteristic of hopping conduction. Notably, the carriers can transit through trap sites in hopping conduction, even though the carrier energy is lower than the barrier energy.[2] To evaluate this behavior, we plotted $V$ versus ln($I$) (Figure S10b), where data points within the gray region exhibited a linear correlation ($R^2 > 0.99$), supporting the hopping conduction model. However, discrepancies from linearity are observed at voltages below 0.9 V, suggesting the involvement of additional conduction mechanisms. Further investigations, including charge transport simulations, are warranted to elucidate the underlying physics beyond experimental observations.

## Table S4: Physical and material parameters used in the 1D compact model.

| Symbol/ Parameter | Value | Symbol/ Parameter | Value |
|---|---|---|---|
| $M_{Ag}$: molecular mass of Ag | 107.87 g mol$^{-1}$ | $\rho_{Ag}$: mass density of Ag | 10490 kg m$^{-3}$ |
| $c_{ion}$: ion concentration | 3 × 10$^{26}$ m$^{-3}$ | $m^*$: effective mass | 0.5 $m_e$ |
| | | $m_e$: electron mass | 9.11 × 10$^{-31}$ kg |
| $r_{ac}$: radius of active electrode reaction area | 10 nm | $r_{fil}$: filament radius | 4 nm |
| $r_{is}$: cross-sectional radius of ion hopping process | 4 nm | $L$: length of switching/insulating layer | 4.3 nm |
| $z$: charge transfer number | 1 | $\alpha_{et}$: electron transfer coefficient | 0.5 |
| $k_{0,et}$: electron transfer reaction rate | 8 × 10$^2$ m s$^{-1}$ | $\Delta G_{et}$: electron transfer activation barrier | 0.58 eV |
| $\Delta W_0$: eff. tunnelling barrier height | 0.3 eV | $R_S$: series resistance | 120 Ω |
| $a$: mean ion hopping distance | 0.25 nm | $f$: ion hopping attempt frequency | 1 × 10$^{14}$ Hz |
| $\Delta G_{hop}$: ion migration barrier | 0.3 eV | $R_{el}$: resistance of electrodes | 40 Ω |
| $\rho_{fil}$: filament's electronic resistivity | 1.59 × 10$^{-8}$ Ω m | $\alpha_{nuc}$: electron transfer coefficient for nucleation | 0.2 |
| $T$: temperature | 298 K | $N_c$: number of atoms to achieve stable nucleus | 1 |
| $t_{0,nuc}$: prefactor of nucleation time | 2 × 10$^{-7}$ s | $\Delta G_{nuc}$: nucleation activation energy | 0.6 eV |

## Table S5: Physical and material parameters used in the 2D continuum model.

| Symbol/ Parameter | Value | Symbol/ Parameter | Value |
|---|---|---|---|
| $M_{Ag}$: molecular mass of Ag | 107.87 g mol$^{-1}$ | $z_c$: charge transfer number | 1 |
| $\rho_{Ag}$: mass density of Ag | 10490 kg m$^{-3}$ | $\rho_{Pd}$: mass density of Pd | 12 020 kg m$^{-3}$ |
| $c_{p,Ag}$: specific heat capacity of Ag | 230 J kg$^{-1}$ K$^{-1}$ | $c_{p,Pd}$: specific heat capacity of Ag | 244 J kg$^{-1}$ K$^{-1}$ |
| $c_{p,MoS2}$: specific heat capacity of MoS$_2$ | 300 J kg$^{-1}$ K$^{-1}$ | $c_{ion}$: ion concentration | 3 × 10$^{26}$ m$^{-3}$ |
| $\sigma_{Ag}$: specific conductivity of Ag | 6.30 × 10$^7$ S m$^{-1}$ | $\sigma_{Pd}$: specific conductivity of Pd | 9.26 × 10$^6$ S m$^{-1}$ |
| $\kappa_{Ag}$: thermal conductivity of Ag | 429 W m$^{-1}$ K$^{-1}$ | $\kappa_{Pd}$: thermal conductivity of Pd | 72 W m$^{-1}$ K$^{-1}$ |
| $\kappa_{MoS2}$: thermal conductivity of MoS$_2$ | 1.1 W m$^{-1}$ K$^{-1}$ | $T_0$: reference temperature | 300 K |
| $\alpha_{e,Ag}$: electrical temperature coefficient 1$^{st}$ degree of Ag | 3.8 × 10$^{-3}$ K$^{-1}$ | $\lambda_e$: electron mean free path | 53.3 nm |
| $\beta_{e,Ag}$: electrical temperature coefficient 2$^{nd}$ degree of Ag | 4 × 10$^{-7}$ K$^{-2}$ | $I_{cc}$: maximum current | 100 µA and 1 mA |
| $\alpha_{t,Ag}$: thermal temperature coefficient 1$^{st}$ degree of Ag | 1.9 × 10$^{-4}$ K$^{-2}$ | $V_{eq}$: equilibrium potential | 0.5 V |
| $\alpha_{e,Pd}$: electrical temperature coefficient 1$^{st}$ degree of Pd | 3.7 × 10$^{-3}$ K$^{-1}$ | $f_{hop}$: hopping frequency | 1 × 10$^{14}$ Hz |
| $\beta_{e,Pd}$: electrical temperature coefficient 2$^{nd}$ degree of Pd | −5.9 × 10$^{-7}$ K$^{-2}$ | $\Delta G_{hop}$: ion migration barrier | 0.15 eV |
| $\alpha_{t,Pd}$: thermal temperature coefficient 1$^{st}$ degree of Pd | 2.9 × 10$^{-4}$ K$^{-1}$ | $m^*$: effective electron mass | 0.5 $m_e$ |
| $k_0$: electron transfer reaction rate | 3.45 × 10$^{-31}$ mol m s$^{-1}$ | $p$: scattering coefficient | 0 |
| $a_{hop}$: mean ionic hopping distance | 0.3 nm | $V_0$: current limitation fitting coefficient | 0.01 V |
| $\Delta G_A$: activation energy | 0.62 eV | $\alpha$: current limitation fitting coefficient | 0.5 |
| $\Delta W_0$: tunnel barrier | 0.3 eV | $r_{tem,exp}$: mechanical stress fitting coefficient | 2 nm to 20 nm |
| $W_{v,0}$: AE dissolution energy barrier coefficient | 0.03 eV | $W_{m,0}$: mechanical stress barrier coefficient | 0.05 eV |

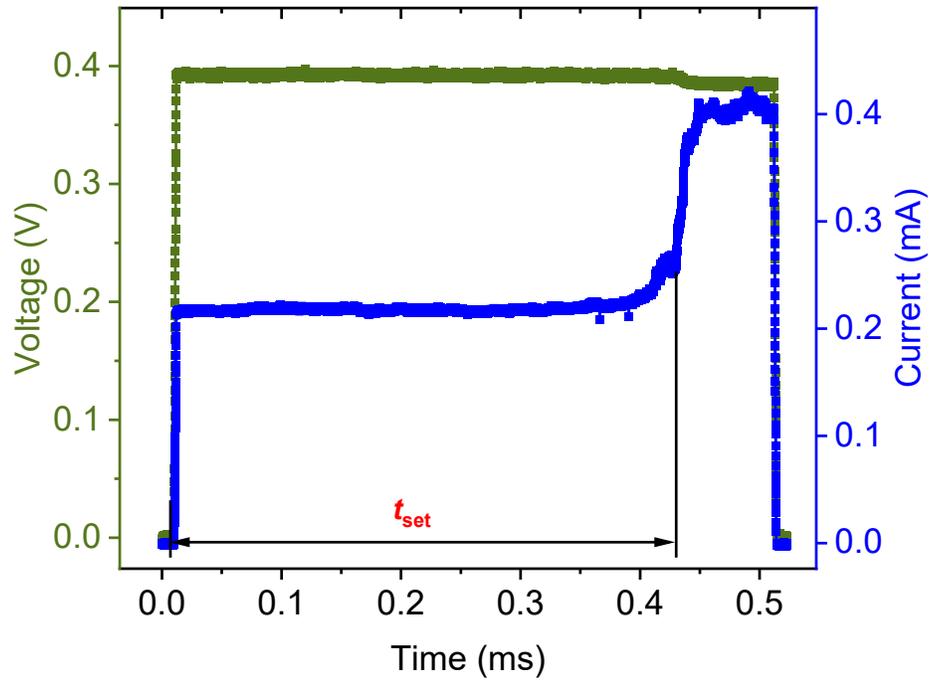

Figure S13: IRS ↔ LRS transition under a 0.4 V single-voltage pulse, exhibiting a $t_{set}$ of ~0.425 ms.

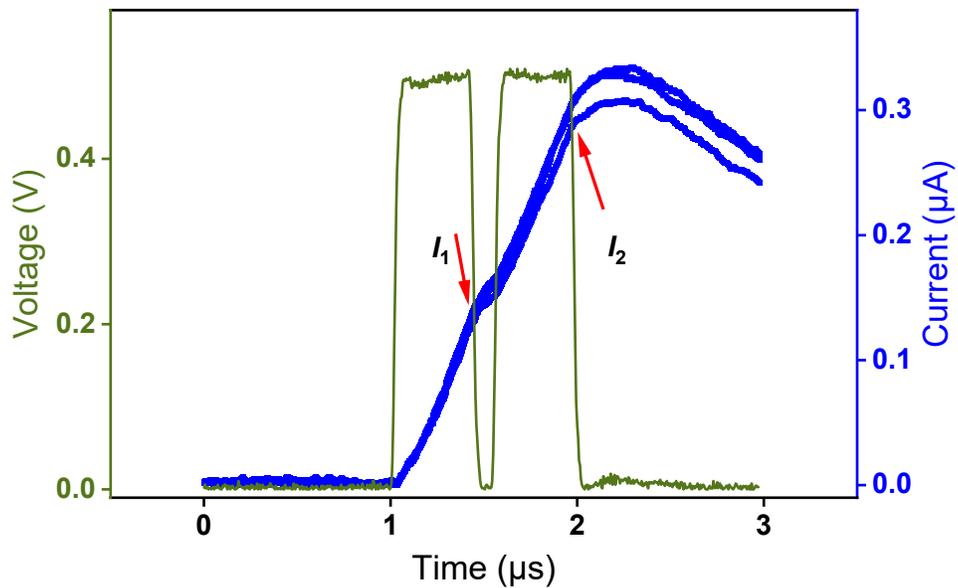

Figure S14: Current response of the memristor for the PPF under two consecutive voltage pulses with a Δ$t$ of 0.1 μs. The current values $I_1$ and $I_2$ were recorded at the end of the first and second pulses, respectively, for each cycle.

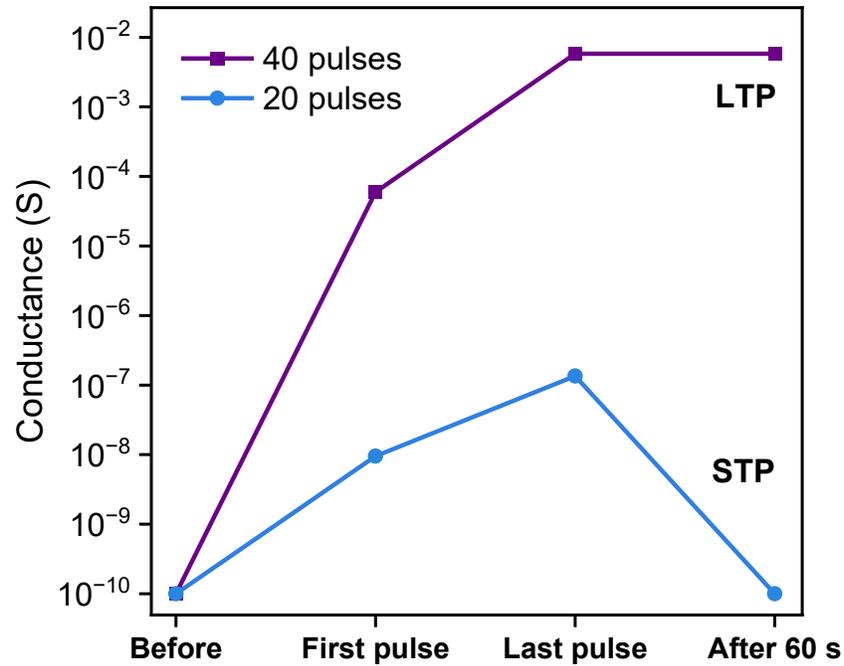

Figure S15: Conductance values extracted before pulse stimulation, at the first and last pulses, and after 60 s, illustrating the transition from STP to LTP.